**Title page**

**Title: Consistency and differences between centrality measures across distinct classes of networks**

Authors: Stuart Oldham [1*], Ben Fulcher [1,2], Linden Parkes [1], Aurina Arnatkevičiūtė [1], Chao Suo [1], Alex Fornito [1]

[1] Brain and Mental Health Research Hub, School of Psychological Sciences and the Monash Institute of Cognitive and Clinical Neurosciences (MICCN), Monash University, Australia.

[2] School of Physics, University of Sydney, Australia.

Conflict of Interest: All authors report no conflict of interest.

**\*Corresponding author:**

Stuart Oldham

Brain and Mental Health Research Hub, School of Psychological Sciences and the Monash Institute of Cognitive and Clinical Neurosciences (MICCN), Monash University.

E: stuart.oldham@monash.edu

The roles of different nodes within a network are often understood through centrality analysis, which aims to quantify the capacity of a node to influence, or be influenced by, other nodes via its connection topology. Many different centrality measures have been proposed, but the degree to which they offer unique information, and such whether it is advantageous to use multiple centrality measures to define node roles, is unclear. Here we calculate correlations between 17 different centrality measures across 212 diverse real-world networks, examine how these correlations relate to variations in network density and global topology, and investigate whether nodes can be clustered into distinct classes according to their centrality profiles. We find that centrality measures are generally positively correlated to each other, the strength of these correlations varies across networks, and network modularity plays a key role in driving these cross-network variations. Data-driven clustering of nodes based on centrality profiles can distinguish different roles, including topological cores of highly central nodes and peripheries of less central nodes. Our findings illustrate how network topology shapes the pattern of correlations between centrality measures and demonstrate how a comparative approach to network centrality can inform the interpretation of nodal roles in complex networks.

# Introduction

Connections are often distributed heterogeneously across the elements of many real-world networks, endowing each node with a specific pattern of connectivity that constrains its role in the system. One popular way of characterizing the role of a node in a network is by using one or more measures of centrality. These measures aim to quantify the capacity of a node to influence, or be influenced by, other system elements by virtue of its connection topology [1–4]. Accordingly, centrality measures are often used to identify highly central or topologically important nodes, commonly referred to as hubs, that play a key role in many diverse kinds of networks. Examples include individuals who enhance the spread of disease in a population [5], proteins that are indispensable for an organism's survival [6], researchers that are frequent collaborators in scientific collaboration networks [7], and brain regions thought to be important for regulating consciousness in functional brain networks [8,9].

Whether a node is ranked highly on a given centrality measure depends on the dynamical processes that are assumed to take place on the network [1]. For instance, nodes that are ranked as highly central according to measures that assume routing of information along shortest paths may not be ranked as highly by measures that assume diffusive dynamics [10,11]. Over 200 centrality measures have been proposed to date [12], each making different assumptions about network dynamics and the topological properties that are important for driving those dynamics. In addition, some centrality measures capture local information (e.g., with respect to immediate network neighbours), whereas others quantify how a node is situated within the global network context [13–15]. In theory, these measures should capture different aspects of network topology, and thus identify different kinds of node roles and, accordingly, different highly-central hub nodes. However, theoretical and conceptual differences between centrality measures do not always translate into empirical differences in real-world networks. For example, two different centrality measures may behave similarly on real-world networks, thus being practically redundant despite their distinct theoretical foundations.

The extent to which different centrality measures offer unique or redundant information depends on the topology of the network (e.g., see Fig. 1). Past empirical work has investigated correlations between the scores assigned by different centrality measures in a number of real-world networks, such as scientific collaboration networks, airline networks, and internet routing networks, finding that the correlations between centrality measures—while typically moderate to high—can vary substantially from one network to another [16,17]. As an example, closeness and eigenvector centrality were very highly correlated in a network of collaborations between high-energy physicists ($r = 0.91$), but not in a power grid network ($r = -0.04$) [17]. The specific reasons for these variations in correlations between centrality measures, hereafter referred to as centrality measure correlations (CMCs), in different networks remains unclear.

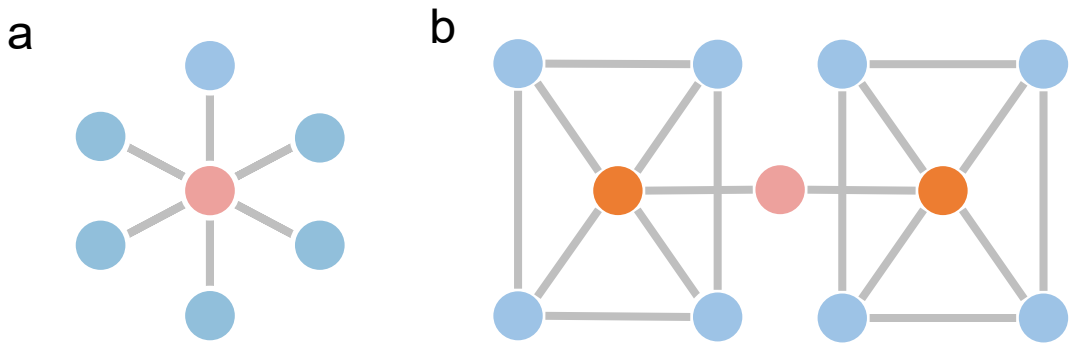

**Figure 1. Basic properties of topological centrality.** (a) shows an example of a star network. The red node has maximal degree (greatest number of connections), closeness (is a short distance from other nodes) and betweenness (lies on many shortest-paths between nodes) in this network. In this case these three centrality measures are perfectly concordant. (b) shows a network in which centrality measures are not concordant. The red node has the highest betweenness and closeness, but it has the lowest degree in the network.

What are the topological properties that influence the CMC structure of a network? Recent theory, developed in the analysis of social networks, has pointed to the *neighbourhood inclusion preorder* of a network as being a major determinant of CMCs (for a more detailed description, see Methods) [18–20]. This property can be quantified using the *majorization gap*, which measures the topological distance of a network from a threshold graph, a type of network in which all centrality measures should rank nodes the same way [18]. Networks that have a low majorization gap, and which are thus topologically similar to a threshold graph, exhibit higher correlations between centrality measures [20]. Another body of work has shown that networks with a large spectral gap, quantified as the difference between the first and second eigenvalues of the adjacency matrix, have very high correlations between centrality measures that quantify walks between nodes [21–23] (for example, subgraph and eigenvector centrality). Clustered, modular networks can reduce CMCs by dissociating measures that quantify centrality within local neighbourhoods of nodes (e.g., degree, leverage) from those indexing centrality across the entire network (e.g., betweenness, closeness). This is because a node may have high local centrality (highly connected with nodes in the same module), but low global centrality (unconnected to nodes in other modules), or vice-versa [24]. Other studies have examined the role of network edge density and the impact of specific node or edge removals on the network [14,25–27].

The extent to which any association between topology and CMC structure generalizes beyond this past work is unclear, as these studies have typically focused only on specific network classes (e.g., social, synthetic), used networks varying within a limited range of sizes and densities, explored just a few types of network organization, or examined a small subset of centrality measures. A systematic evaluation of CMCs, quantified across a broad array of centrality metrics and in a large set of different classes of networks, has not been performed. Furthermore, given the abundance of centrality measures proposed, many of which are highly correlated to each other when applied to real-world networks, it is important to understand whether there are benefits to using multiple centrality measures, or whether there is a reduced, canonical set of measures for capturing nodal roles in most applications. Past research has found that using multiple centrality measures to define multivariate profiles can offer a better description of nodal roles in the network [28,29]. Broad, comparative studies—such as those performed recently for time-series analysis [30]—allow us to uncover empirical relationships between the large and interdisciplinary literature on centrality measures for network data. Where the selection of which centrality measure to apply to a given network analysis task is typically done subjectively, the combination of many centrality measures together can offer a more systematic and comprehensive framework in which the most useful measures can be informed more objectively from the empirical structure of a given network.

In this article, we evaluate 17 different centrality measures across 212 networks. We examine how CMCs vary across the networks and characterize the association between global topological properties of each network and CMC variation. We also examine how multivariate profiling of nodal centrality can be used to gain insight into the roles that different nodes play a given network.

**Methods**

**Centrality Measures.** We used 17 different centrality measures to analyse each network, focusing on centrality measures that are commonly used in the network science literature, or which have received recent interest. Each measure used is listed in Table 1; definitions and further details are in the Supplementary Information. Analysis was performed in MATLAB 2017a. The code for all centrality measures were either obtained from the Brain Connectivity Toolbox (BCT) [31], MatlabBGL library, or were written in custom code, available at [https://github.com/BMHLab/CentralityConsistency]. All data generated or analysed in the current study are available in the figshare repository, [https://figshare.com/s/22c5b72b574351d03edf].

Centrality measures are often defined in relation to the different ways in which information is thought to propagate across nodes, which can occur through: (1) *walks*, which follow an unrestricted trajectory through the network; (2) *trails*, which can return to a visited node but cannot reuse an edge; and (3) *paths*, which cannot visit a node or edge more than once [1]. Thus, paths are a subset of trails which, in turn, are a subset of walks. We sought to include measures based on these different propagation approaches, although most centrality measures developed to date have focused on walks and paths.

While not typically thought of as a centrality index, the *participation coefficient* was also included in our set of centrality measures for comparison, as it is frequently used as a measure of nodal roles in networks with community structure [4,24]. The participation coefficient quantifies the distribution of a node's connections across different topological modules of the network, where the modules are defined using a specific community detection algorithm (for a review of community detection algorithms see [32]). The participation coefficient was first introduced to distinguish between different types of network hubs [24] and has been proposed as a singular measure for defining hubs in some classes of networks, such as those based on correlations [33].

| Centrality name | Characteristics of a central node | Equation |
|---|---|---|
| Degree (DC) | Connected to many other nodes [3] | $DC_i = d_i = \sum_{j \neq i} A_{ij}$ |
| Eigenvector (EC) | Connected to many other nodes and/or to other high-degree nodes [34] | $EC_i = \frac{1}{\lambda_1} \sum_j A_{ji} v_j$ |
| Katz (KC) | Connected to many other nodes and/or connected to other high-degree [35] | $KC_i = \alpha \sum_j A_{ji} v_j + \beta$ |
| PageRank (PR) | Connected to many other nodes and connected to other high-degree nodes [36] | $PR_i = \alpha \sum_j A_{ji} \frac{v_j}{k_j} + \beta$ |
| Leverage (LC) | Has a higher degree than its neighbours [37] | $LC_i = \frac{1}{d_i} \sum_{j \in h(i)} \frac{d_i - d_j}{d_i + d_j}$ |
| H-index (HC) | Connected to many other high-degree nodes [38] | $HC_i = \max_{1 \leq h \leq d_i} \min(|\mathcal{N}_{\geq h}(i)|, h)$ |
| Laplacian (LAPC) | Removal of this node would greatly impair the network [39,40] | $LAPC_i = d_i^2 + d_i + 2 \sum_{j \in \mathcal{N}(i)} d_j$ |
| Shortest-path Closeness (CC) | Low average shortest path length to other nodes in the network [41] | $CC_i = \frac{N}{\sum_j l_{ij}}$ |
| Subgraph (SC) | Involved in many closed short-range walks [42] | $SC_i = [e^A]_{ii}$ |
| Participation coefficient (PC) | Connections distributed across different topological modules [24] | $PC_i = 1 - \sum_{m=1}^{M} \left(\frac{d_i(m)}{d(i)}\right)^2$ |
| Total Communicability (TCC) | Can be easily reached by a walk from any other node [21] | $TCC_i = \sum_j [e^A]_{ji}$ |
| Random-walk Closeness (RWCC) | Can be easily reached by a random-walk from any other node [43,44] | $RWCC_i = \frac{N}{\sum_j H_{ji}}$ |
| Information (IC) | Can be easily reached by paths from other nodes [45] | $IC_i = \left(C_{ii} + \frac{\sum_j C_{jj} - 2 \sum_j C_{ij}}{N}\right)^{-1}$ |
| Shortest-path Betweenness (BC) | Lies on many shortest topological paths linking other node pairs [3] | $BC_i = \sum_{p \neq i, p \neq q, q \neq i} \frac{g_{pq}(i)}{g_{pq}}$ |
| Communicability betweenness (CBC) | Takes part in many walks between pairs of other nodes [46] | $CBC_i = \frac{1}{\acute{C}} \sum_p \sum_q \frac{G_{piq}}{G_{pq}}, p \neq q, q \neq i$ |
| Random-walk Betweenness (RWBC) | Takes part in many random walks between pairs of other nodes [47] | $RWBC_i = \frac{\sum_{p<q} I_i^{(pq)}}{\frac{1}{2} N(N-1)}$ |
| Bridging (BridC) | Forms key links between high degree nodes [48] | $BridC_i = BC_i \times Bc_i$ |

**Table 1. Definitions for Centrality Measures.** $A$ = adjacency matrix; $d_i$ = degree of node $i$; $\lambda_1$ = leading eigenvalue of $A$; $v$ = leading eigenvector of $A$; $\alpha$ = penalty on distant connections to a node's centrality score; $\beta$ = preassigned centrality constant $h(i)$ = the neighbours of node $i$; $\mathcal{N}_{\geq h}(i)$ = neighbours of node $i$ which have at least a degree of $h$; $N$ = number of nodes in a network; $l_{ij}$ = length of the shortest between nodes $i$ and $j$; $e^A$ = matrix exponential of $A$; $M$ = number of modules in a network; $d_i(m)$ = neighbours of node $i$ which are part of module $m$; $H$ = the matrix of mean-first passage times between nodes in a network; $C = (L + J)^{-1}$ where $L$ is the Laplacian of $A$ and $J$ is a $N \times N$ matrix with all elements equal to one; $g_{pq}$ = the number of shortest-paths between nodes $p$ and $q$; $g_{pq}(i)$ = the number of shortest-paths between nodes $p$ and $q$ which pass through $i$; $G_{pq}$ = number of walks between nodes $p$ and $q$; $G_{piq}$ = number of walks between nodes $p$ and $q$ involving node $i$; $\acute{C} = (N-1)^2 - (N-1)$ which is a normalisation term; $I_i^{(pq)}$ = current flowing through nodes $p$ and $q$ which passes through node $i$; $Bc_i = d_i^{-1} / \sum_{j \in \mathcal{N}(i)} d_j^{-1}$. All measures here are defined for unweighted networks, see Supplementary Information for information on weighted versions.

**Network Data.** Nearly all networks were obtained from freely-available sources. We examined 107 networks compiled by Ghasemian and colleagues [49] from the Index of Complex Networks (ICON) [50], together with a further 104 networks sourced by searching ICON for networks of varying sizes and domains. An additional network, the human structural brain network, was generated from diffusion-weighted magnetic resonance imaging data from the Human Connectome Project [51] (see Supplementary Information for details). we thus considered a total of 212 networks. Each network, comprising $N$ nodes and $E$ edges, was represented as an $N \times N$ adjacency matrix. For the main analysis, each network was treated as unweighted (any edge weight information was removed) and undirected (any unidirectional edges were made bi-directional). Additionally, if the network comprised multiple components, only the largest connected component was considered. In addition, weighted analysis was performed for 39 networks with edge-weight information.

To examine the extent to which simple network properties—such as number of nodes, edges, and degree/strength distribution—contribute to the CMCs for a network, we compared the empirical networks to a set of matched surrogate networks. For each empirical network, we generated 100 unconstrained and 100 constrained surrogate networks. Unconstrained surrogate networks were created using a variant of the Erdős-Rényi generative model [52] which guaranteed the network would be non-fragmented, while preserving the number of nodes, number of edges, and the distribution of edge weights of the original network. Constrained surrogate networks were generated using the Maslov-Sneppen algorithm [53] for unweighted networks and a modified version for weighted networks [31]. The constrained surrogates preserve the number of nodes and edges, in addition to the degree sequence and approximate node strength (weighted degree) distributions. See Supplementary Information for more on the surrogate generation algorithms. Due to the computational complexity of calculating random-walk betweenness centrality and communicability betweenness centrality, we did not compute these measures for the surrogate networks.

**Centrality Measure Correlations (CMCs).** We used Spearman's $\rho$ to calculate the correlation between the nodal scores assigned by any two centrality measures. This statistic was used to quantify CMCs because many such relationships were nonlinear yet almost always monotonic, and many centrality metrics have a non-Gaussian distribution [20]. CMCs were computed in every network for all pairs of centrality metrics. To find which centrality measures were consistently highly correlated across networks (indicating redundancy), we took the mean CMC for each pair of metrics across all networks, which we term the *mean between-network CMC*. We also quantified the variability of CMCs across networks as the *between-network CMC standard deviation*.

**Assessing the Relationship Between Network Topology and CMCs.** Given the assumed relationship between network topology and CMCs (e.g., Fig. 1), we examined how CMCs vary as a function of eight different global network properties: connection density, assortativity, clustering, connection density, global efficiency, diffusion efficiency, modularity, majorization gap, and spectral gap. Further details on how these global topological properties were calculated can be found in the Supplementary Information. Briefly, the connection density of a network, κ, is the proportion of connections that are present in a network relative to the total number of possible connections. Previous work has shown that networks with higher density show higher CMCs [27]. In the limit of $\kappa = 1$, the network is fully connected and all nodes are identical. As the density decreases, there is more variability in how the connections in the network can be arranged, and this is likely to result in centrality measures diverging and thus becoming less correlated.

Assortativity, clustering and global efficiency are commonly used descriptors of global network topology. Assortativity measures the extent to which nodes preferentially connect to other nodes with similar degree [54]. Clustering measures the proportion of closed triangles present in the network, and is often taken as a measure of cliquish connectivity [55]. Global efficiency is inversely related to the characteristic path length of a network, and is thus a useful descriptor for networks characterized by shortest-path routing [56]. Diffusion efficiency is an analogous measures that captures the efficiency of a network in supporting communication governed by a diffusion process [11].

Modularity is the extent to which a network contains groups of nodes that are densely interconnected with each other but sparsely connected to nodes outside the group [54]. Prior work has indicated that networks with stronger modularity show weaker CMCs [24]. Modules can enhance topological heterogeneity in a network, dissociating centrality metrics that favour high within-module connectivity (high local neighbourhood connectivity) from high between-module connectivity (globally integrative connectivity). We quantified modularity using the widely-used $Q$ metric (Newman and Girvan, 2004), and modules were identified using the Louvain algorithm [58] combined with a consensus clustering procedure (50 iterations with $\tau = 0.4$) [59,60] to address algorithmic degeneracy [61] (see Supplementary Information).

The majorization gap quantifies the distance between an empirical network and an idealized network, called a threshold graph [20]. Threshold graphs are formed by adding nodes to a network, one at a time, such that the new node either connects to all existing nodes or connects to no other nodes (see Fig. S1 for an example). Threshold graphs preserve a property known as the neighbourhood-inclusion preorder, which is argued to form the basis of centrality rankings [18,19]. If the neighbours of node $j$ are a subset of the neighbors of node $i$, then node $i$ is said to dominate node $j$, and must have a greater or equivalent level of centrality. The neighbourhood inclusion preorder is the rank ordering of nodes in terms of these dominance relationships, such that nodes that are not dominated by any others are ranked first and are thus more central. Nodes that are dominated by many others are ranked last, and are thus least central (e.g., Fig. S2). As this preorder is complete in threshold graphs, i.e., each node is either dominated by another or not, the centrality rankings of all nodes across different measures in these networks is perfectly concordant. Thus, networks with a larger majorization gap will be more topologically distant from a threshold graph and should have lower CMCs.

The final property investigated was the spectral gap. This property quantifies the quality of a network's 'expansion properties'; namely, whether a network is simultaneously sparse and well-connected. Good expansion networks lack bottlenecks — nodes/edges that, if removed, will fragment the network. A larger the spectral gap has been associated with the correlations between walk-based centrality measures [21–23].

To combine the overall similarity of all pairs of centrality measures into a single value for a network, we took the mean of every CMC within each network to obtain the *mean within-network CMC*. A higher mean within-network CMC indicates that, on average, centrality measures are highly correlated in a network. This value was then correlated with each global topological descriptor. To determine which specific topological descriptor was the best predictor of variations in mean CMC across networks, we used multiple linear regression. In secondary analyses, we examined whether specific CMCs correlated with variations in global topology across networks.

As simple network properties like edge density and the degree/strength distribution can account for many higher-order features network topology, we compared the CMCs of empirical networks to matched surrogate networks. The unconstrained model can be used to determine whether the relationship is explained simply by variations in size and density across networks, while the constrained surrogates can be used to examine the impact of degree sequence and strength distribution in driving this relationship. To allow comparison between different networks and their associated surrogates, we calculated the difference of the empirical networks properties/mean within-network CMCs compared to the mean value obtained in each of the surrogates.

**Clustering nodes based on their centrality profiles.** Finally, we investigated whether combining multiple centrality measures into a multivariate 'centrality profile' for each node could be used to meaningfully cluster nodes into groups with distinct topological roles. Centrality scores were converted to ranks and hierarchical clustering was performed using Ward's minimum variance method [62] for Euclidean distances between pairs of ranked centrality metrics. For visualization, the Davies-Bouldin (DB) index [63] was used to determine a specific resolution to cut the dendrogram and investigate the resulting clusters. The DB index is a ratio of intra-cluster similarity to inter-cluster differences for a given clustering solution; lower values of the DB indicate a better clustering solution. We note that there are many different algorithms for clustering data (including alternative heuristics for forming clusters from a dendrogram) and for dendrogram cutting [64]. Our goal is not to determine any particular clustering solution or approach as robust or optimal, but rather to demonstrate how clustering of centrality profiles may aid in identifying subsets of nodes with distinct topological roles. A forced-directed algorithm was used to visualize node roles in the context of the broader topology of the network [65].

## Results

### Correlations between Centrality Measures.

First, to examine the similarity of centrality measures across different networks, we calculated Spearman correlations between each of the 17 measures listed in Table 1 across each of the 212 networks. All 212 networks were analysed in unweighted form. A separate weighted centrality analysis was performed for 39 of these networks with edge-weight information.

Figure 2 shows the distribution of CMCs of five example unweighted and weighted networks. The distributions of CMCs for all networks are shown in Figure S3. These results indicate that, despite a general trend for most networks to have high and mostly positive CMCs, there is considerable heterogeneity in CMC patterns across different networks, as previously reported [16,17]. This variability did not clearly map track the natural class of the network (i.e., whether the network is social, biological technological, etc; Fig S3).

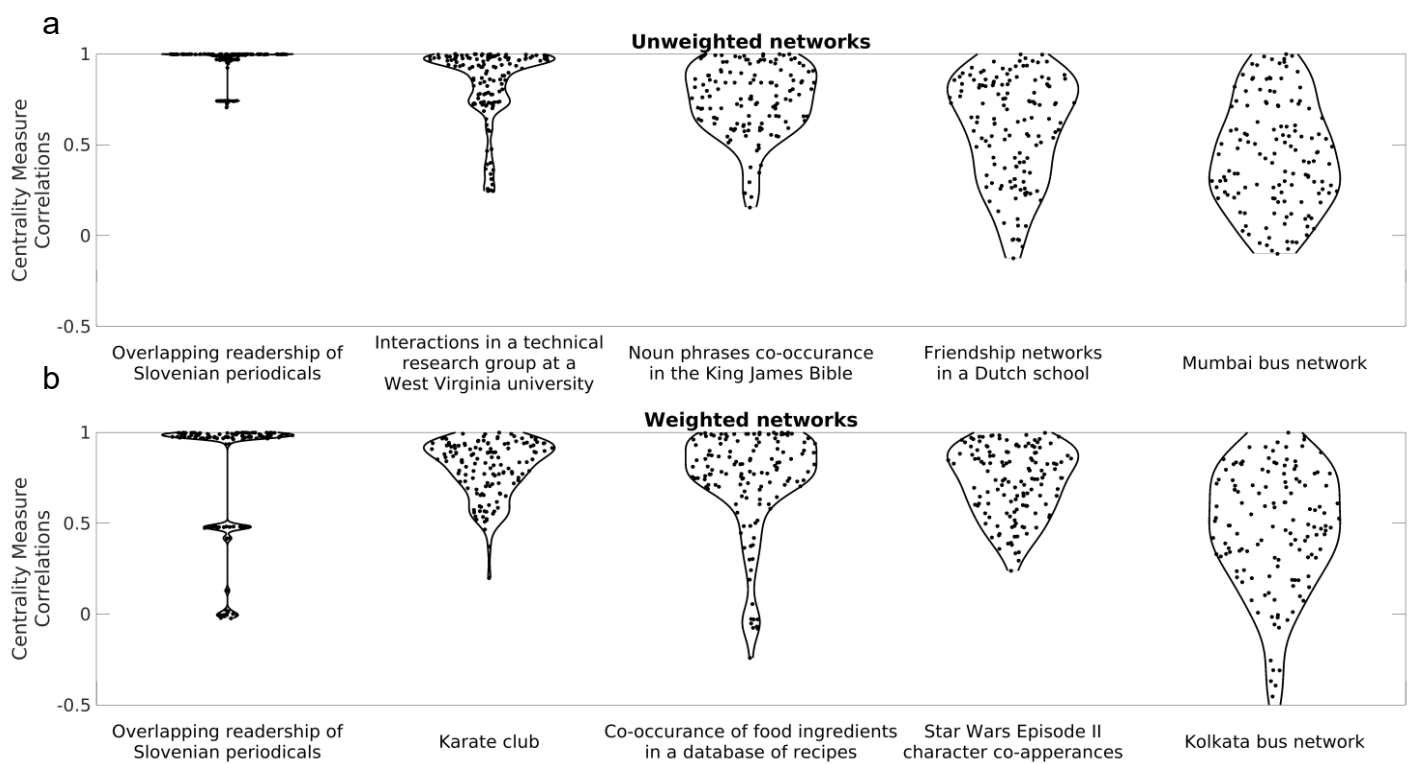

**Figure 2. Distributions of Centrality Measure Correlations (CMCs) for example unweighted and weighted networks.** Distributions of CMCs for every pair of centrality measures for five example (a) unweighted; and (b) weighted networks. Networks have been ordered from highest (left) to lowest (right) median CMC.

To determine which pairs of centrality measures were consistently correlated across networks, we calculated the mean between-network CMC (the mean CMC for each pair of measures across all networks) and standard deviation (standard deviation of CMCs across networks) for each pair of metrics in unweighted (Figs. 3a, 3c) and weighted (Figs. 3b, 3d) networks. Most measures show moderate-to-high correlations across all networks, with 97% of all mean CMCs exceeded 0.5 in unweighted networks and 80% in weighted networks. Weighted CMCs were slightly weaker than their unweighted counterparts. For the 39 networks with edge weight information, we compared the unweighted and weighted centrality measures, finding that both were highly correlated (Fig. S4 and Fig. S5). See supplementary results for further discussion of these findings.

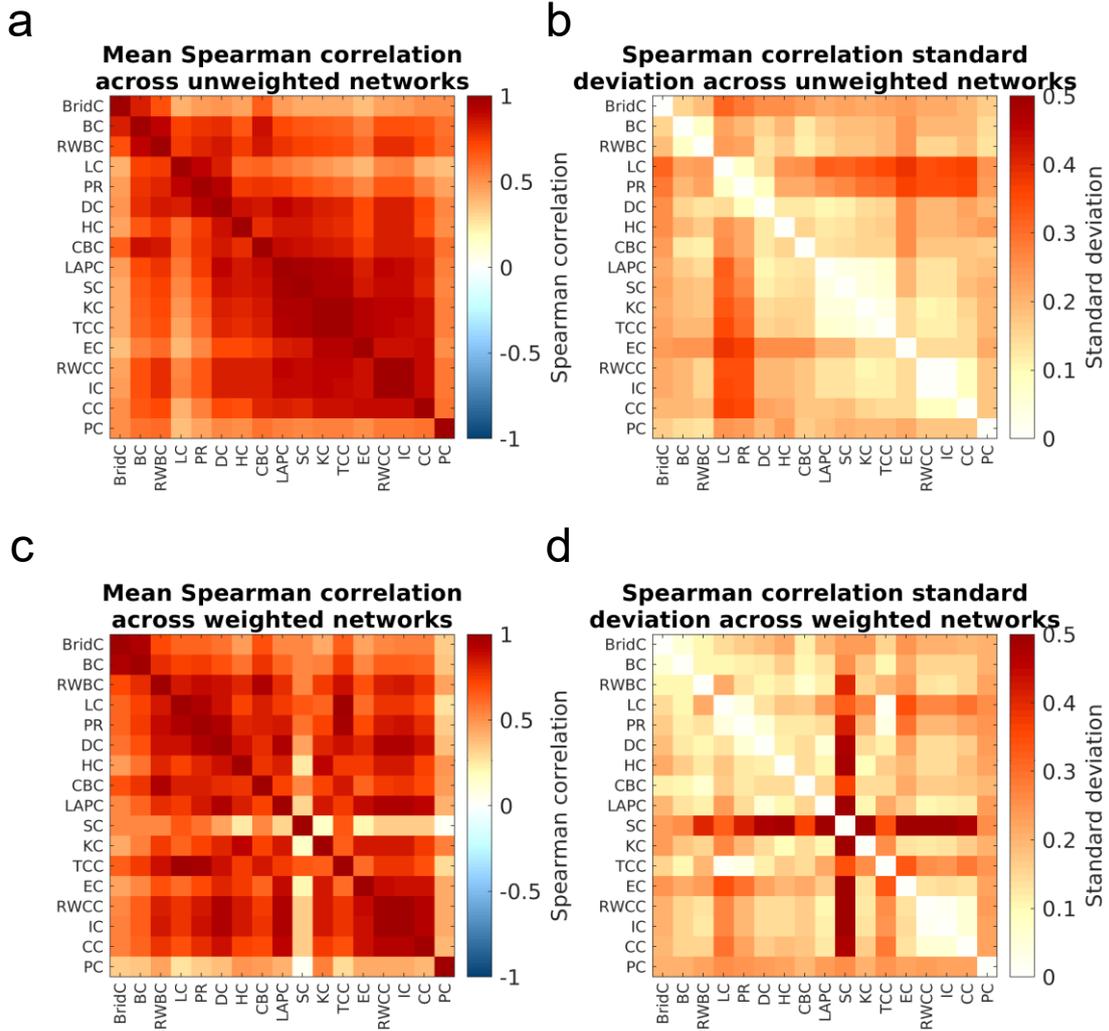

**Figure 3. Mean and standard deviation of between-network CMCs.** (a) and (b) show the between-network CMC mean and standard deviation for unweighted measures, respectively. (c) and (d) show the between-network CMCs mean and standard deviation for weighted measures, respectively.

Several pairs of centrality measures displayed notable relationships. First, random-walk closeness centrality (RWCC) and information centrality (IC) were very highly correlated across networks (ranging from 0.88-1 with a mean correlation of 0.998 in unweighted networks and ranging from 0.937-1 with a mean correlation of 0.996 in weighted networks. Thus, these two theoretically-related measures [66] are practically redundant in most real-world scenarios. Other pairs, like Katz centrality (KC) and total communicability centrality (TCC), were also highly correlated across the wide range of unweighted networks analysed (all $\rho > 0.98$). The participation coefficient and bridging centrality generally had the lowest average correlation with other measures, likely because they are conceptually distinct, and in the case of the participation coefficient, depend on a modular decomposition of the network. Subgraph centrality in weighted networks showed low correlations with other measures, suggesting it may be capturing a unique aspect of node centrality.

**Network Topology and CMCs.**

We now examine how variations in CMCs across different networks relate to differences in the global topological properties of those networks. Specifically, we consider how the mean within-network CMC (the average of all pairwise CMCs within a network) relates to the following eight global network properties: connection density, assortativity, clustering, global efficiency, diffusion efficiency, modularity, majorization gap, and spectral gap.

In unweighted networks, higher mean within-network CMC was correlated with lower values of assortativity, majorization gap, and modularity, and higher values of clustering, density, diffusion efficiency, global efficiency, and spectral gap (Fig. 4). Similar results were obtained for weighted networks (Fig. S6), with some exceptions. First, the correlation between global efficiency and mean within-network CMC was among the strongest for unweighted networks but among the weakest for weighted networks. Conversely, the correlation between assortativity and mean within-network CMC was strong for weighted networks, but weak for unweighted networks. Weighted clustering showed no relationship with CMCs once outliers were removed. Post-hoc analyses indicated that many individual pairs of CMCs correlated with network properties, showing that the relationship between network properties and mean CMCs is representative of a general trend across most pairs of centrality measures, and not driven by a small subset of CMCs (Fig. S7 for unweighted and Fig. S8 for weighted). However, CMCs involving bridging centrality or the participation coefficient had weak correlations with nearly all global properties in both unweighted and weighted networks, further suggesting that these measures may capture a unique aspect of nodal centrality.

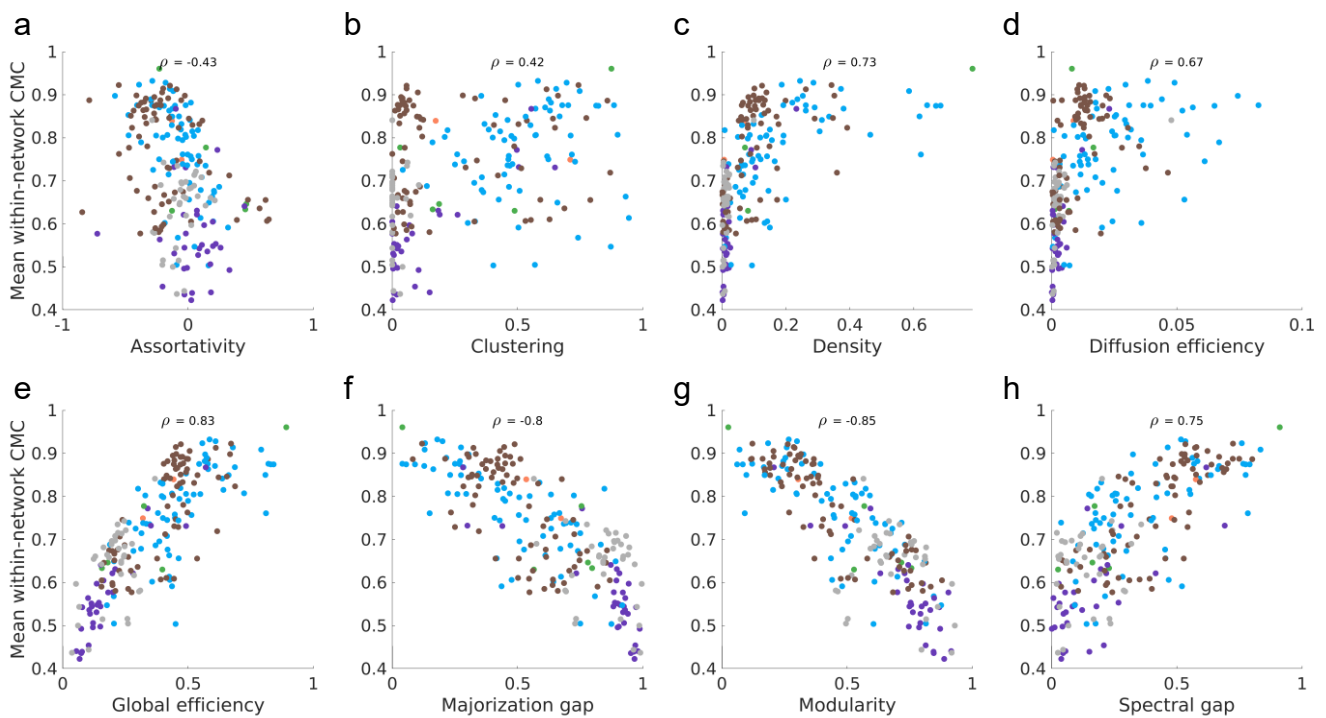

**Figure 4. Association between mean within-network CMC and network properties in unweighted networks.** The association between the mean within-network CMC (the average of all CMCs within a single network) and each of the global topological properties. Networks are coloured by their natural category (blue = social, grey = technological, brown = biological, orange = informational, purple = transportation; green = economic).

We used multiple linear regression to quantify the unique contributions of each topological descriptor to CMC variability across networks (note: network density and diffusion efficiency were excluded due to strong non-linear associations with CMC.) In unweighted networks, modularity was the only significant predictor of mean within-network CMCs (Table S1). As modularity and the majorization gap were highly correlated (Fig. S9), we reran the model excluding one of these properties each time, and found that only modularity was a significant predictor of network CMCs (Table S1). In weighted networks, weighted assortativity explained the most variance in network CMCs. Due to collinearity, modularity and majorization gap were included in separate models. Both were significant predictors in these models, with the former accounting for slightly less variance than the latter (49% vs 55%) (Table S2).

To ensure that the associations between the mean within-network CMC and global topology could not be explained by lower-order features (e.g., density of the network or degree sequence), we examined these associations in surrogate networks matched for number of nodes, number of edges, edge weight distribution (unconstrained surrogate), and degree sequence and strength distribution (constrained surrogate). We compared the mean within-network CMCs and each network property in empirical networks to those obtained in the surrogates. Specifically, we calculated the difference between the mean within-network CMC /network property in the empirical network and the corresponding mean values of the surrogates. A difference greater than zero means the property was higher in the empirical network than the surrogates; conversely, if it was less than zero it was higher in the surrogate networks. A difference close to zero indicates the property is simply a side-effect of the network's density (for unconstrained surrogates) or degree/strength distribution (for constrained surrogates). These results are shown in Figures 5 and 6 for unweighted network while results for weighted networks surrogates are presented in supplementary Figures S10 and S11.

There are three major results from this comparison to the surrogates. First, for most networks, the mean within-network CMC of the surrogate networks (both constrained and unconstrained) was higher or equivalent to the respective matched empirical network (Figs. 5 and 6). Second, unconstrained surrogates also had a higher majorization gap than the empirical networks. Finally, despite the empirical networks and constrained surrogates having the exact same majorization gap (due to the majorization gap being solely determined by the degree sequence of a network), empirical networks often had lower CMCs. Together, these results counter theoretical expectations that a higher majorization gap should be associated with lower CMCs. We discuss potential reasons for this discrepancy between below (and in Supplementary results).

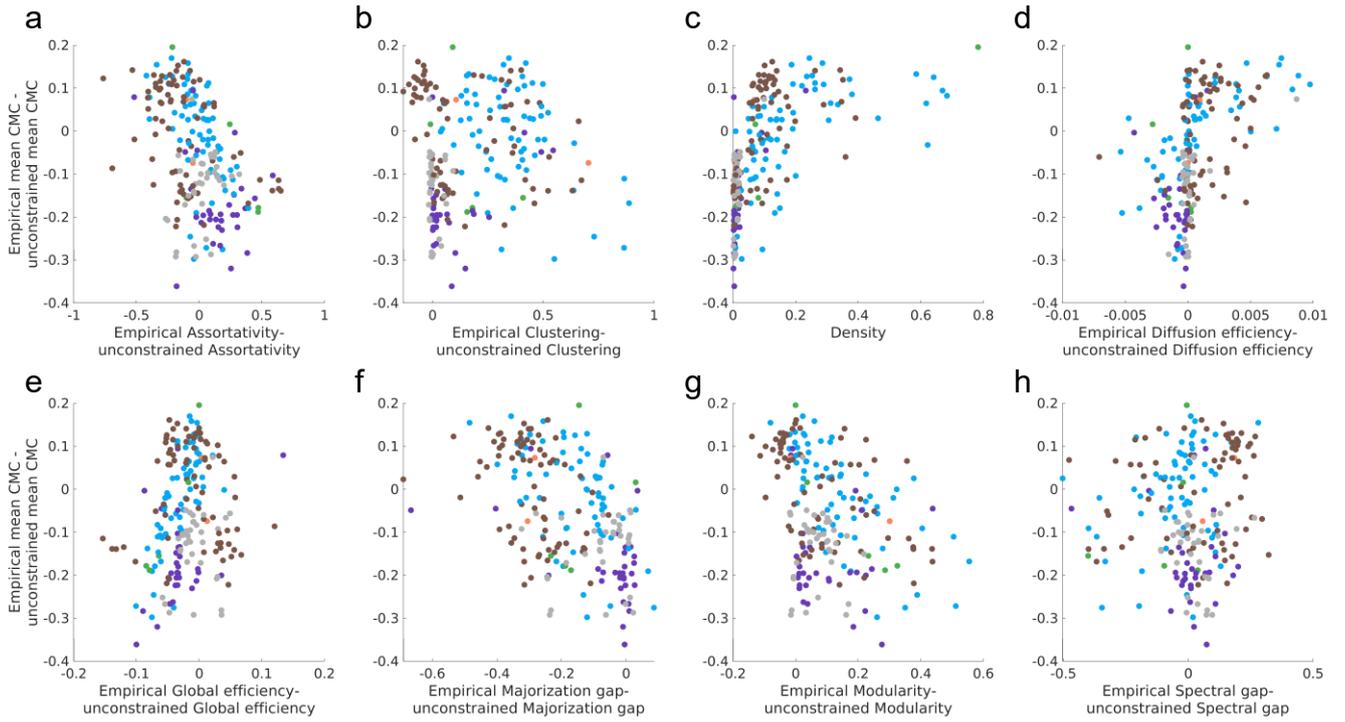

**Figure 5. Difference between unweighted empirical and unconstrained surrogates in mean within-network CMC and network properties.** The y-axis of each plot shows the difference between the empirical networks and unconstrained surrogates mean within-network CMC. The x-axis shows the difference between the empirical networks and unconstrained surrogates on a particular property (except for (c) as the unconstrained surrogates have the same density as the empirical network). On both axis, except for the x-axis in (c), a negative value indicates the empirical network had a lower value than the mean value of the surrogates, while a positive value indicates the empirical networks had a larger value. Points are coloured by the natural category of the empirical network (blue = social, grey = technological, brown = biological, orange = informational, purple = transportation; green = economic).

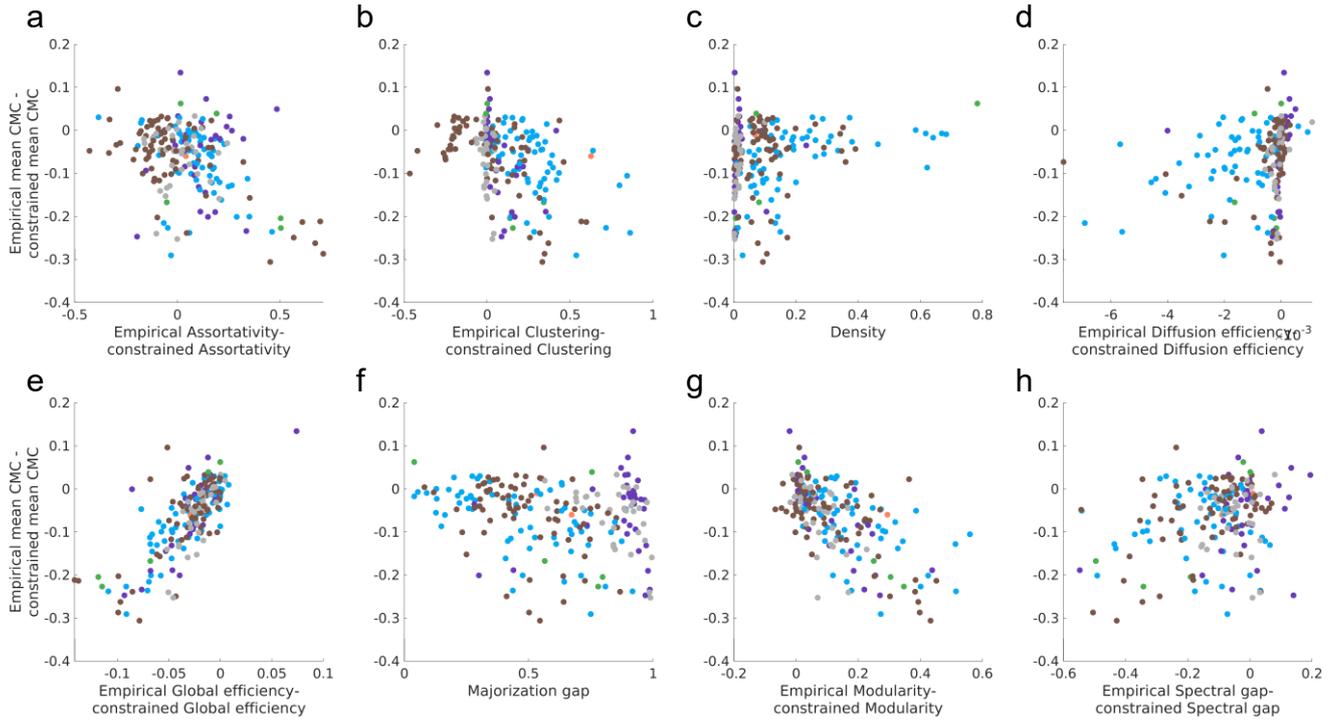

**Figure 6. Difference between unweighted empirical and constrained surrogates in mean within-network CMC and network properties.** The y-axis of each plot shows the difference between the empirical networks and constrained surrogates mean within-network CMC. The x-axis shows the difference between the empirical networks and constrained surrogates on a particular property (except for (c) and (f) as the constrained surrogates have the same density and majorization gap as the empirical network). On both axis, except for the x-axis in (c) and (f), a negative value indicates the empirical network had a lower value than the mean value of the surrogates, while a positive value indicates the empirical networks had a larger value. Points are coloured by the natural category of the empirical network (blue = social, grey = technological, brown = biological, orange = informational, purple = transportation; green = economic).

**Centrality-Based Clustering of Nodes.**

We now use hierarchical clustering to investigate whether multiple centrality measures can be used in combination to identify distinct roles for nodes. Due to the consistent high correlations ($\rho > 0.99$) between random-walk closeness and information centrality, we excluded random-walk closeness from this analysis.

In most networks, the Davies-Bouldin (DB) criterion, a measure of the quality of a given clustering solution, suggested a two-cluster solution. Nearly all networks contained a subset of nodes with high scores across most measures, and another subset with low scores across most measures. The two-cluster solution often favoured one of these groups, such that either all nodes with low centrality were grouped in one cluster and the remaining nodes in the other (e.g., Fig. 7), or vice-versa (e.g., Fig. 8). Such subsets were also apparent went examining finer-grained clustering solutions.

While a putative core of high-scoring nodes and a periphery of low-scoring nodes was consistently found across nodes and clustering resolutions, distinct patterns were found for nodes interposed between these two subsets across different networks. Broadly these patterns can be classified into two types, characterized by either (a) a gradual progression from high-scoring core nodes to low-scoring periphery nodes (Fig. 8A, see also Figs S15-S17), or (b) a semi-discrete cluster structure observable at different resolutions (Fig. 7A, see also Figs S12-S14), in which each cluster has a distinctive profile of scores across different centrality measures. An example of one such intermediate cluster present in several networks comprises nodes that score highly on closeness (e.g., shortest-path closeness, total communicability, subgraph, information) and eigenvector-like (e.g., eigenvector, Katz) measures of centrality, but low on betweenness-based (shortest-path, random-walk, communicability) measures (e.g. Fig. 7 blue cluster; Fig. S13 purple cluster). These nodes were thus topologically positioned within a central core of the network (accounting for their high closeness), were connected to other nodes with high degree (accounting for their high eigenvector values) yet lacked connections to nodes outside of the main cluster (thus having low betweenness and participation coefficient scores). Other intermediate clusters varied depending on the network, and may thus define nodes serving unique roles within each specific system.

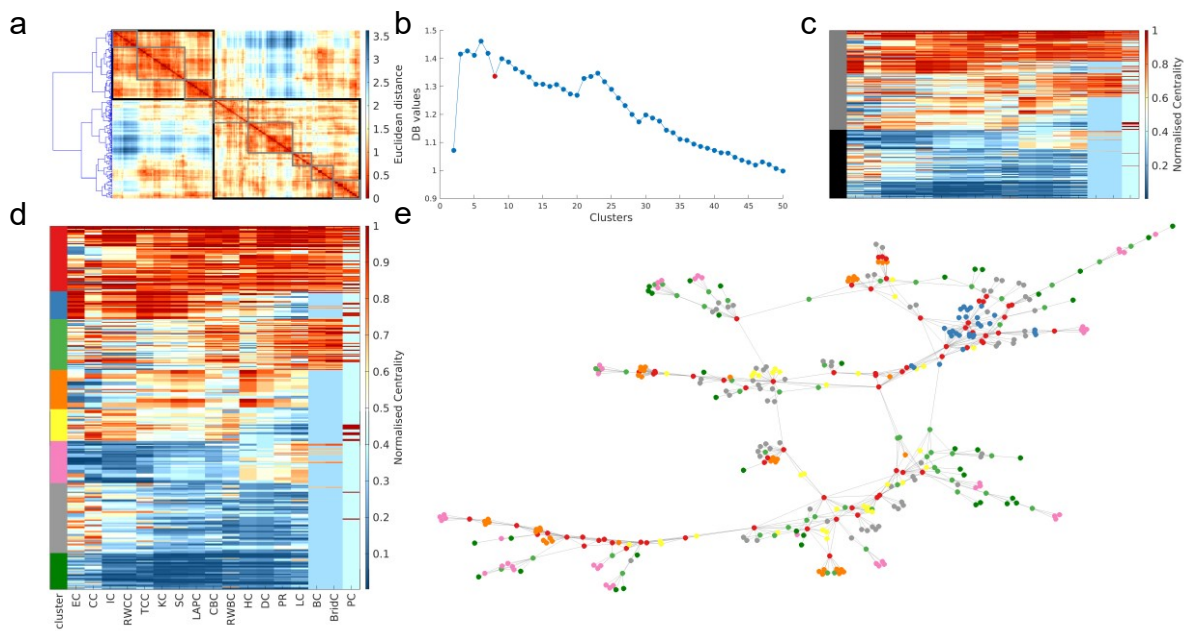

**Figure 7. Multivariate centrality profiling of the network science author collaboration network.** (a) shows the dendrogram projected alongside the distance matrix of node pairs (ranks scores were normalised to be in the range 0-1 with 1 indicating the highest rank). The black and grey boxes and indicate the clusters when a two-cluster and eight-cluster solution is used, respectively. (b) displays the results for the Davies-Bouldin (DB) criterion. A lower DB value represents a better clustering solution. The solution shown in (d) and (e) is labelled in red. Only the first 50 clustering solutions are shown for ease of visibility. (c) shows the matrix of nodal centrality scores (each row is a node and each column is a measure) and how these are clustered in a two-cluster solution (the black and grey represent the two different clusters). (d) shows the matrix of nodal centrality scores as well as the clusters each node was assigned to. (e) shows a topological representation of the network, produced using the force-directed layout algorithm, where each node is coloured according to the cluster it was allocated to in (d).

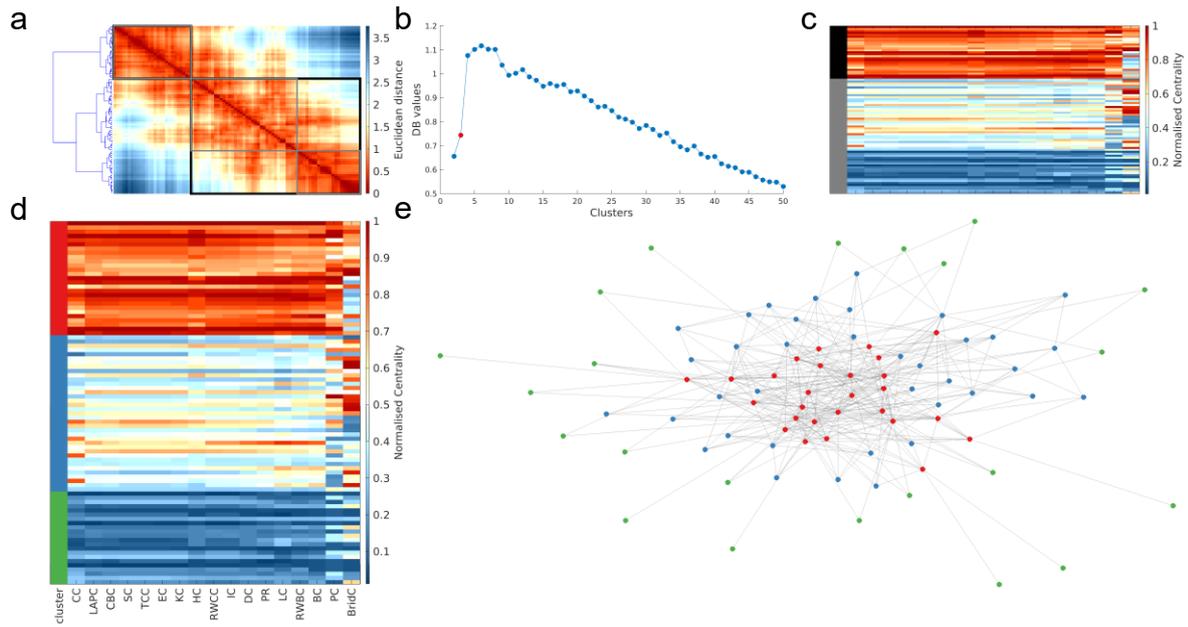

**Figure 8. Multivariate centrality profiling of trophic-level species interactions in a New Zealand stream.** (a) shows the dendrogram projected alongside the distance matrix of node pairs (ranks scores were normalised to be in the range 0-1 with 1 indicating the highest rank). The black and grey boxes and indicate the clusters when a two-cluster and three-cluster solution is used, respectively. (b) displays the results for the Davies-Bouldin (DB) criterion. A lower DB value represents a better clustering solution. The solution shown in (d) and (e) is labelled in red. Only the first 50 clustering solutions are shown for ease of visibility. (c) shows the matrix of nodal centrality scores (each row is a node and each column is a measure) and how these are clustered in a two-cluster solution (the black and grey represent the two different clusters). (d) shows the matrix of nodal centrality scores as well as the clusters each node was assigned to. (e) shows a topological representation of the network, produced using the force-directed layout algorithm, where each node is coloured according to the cluster it was allocated to in (d).

**Discussion.**

We evaluated CMCs between 17 different centrality measures in 212 networks to determine how variation in the strength of CMCs across networks tracks differences in global topological properties. We also investigated whether subsets of nodes with consistent topological roles, including network hubs, could be identified based on their multivariate centrality profiles. We find that centrality measures show moderate-to-high positive correlations across most networks; modularity is the strongest predictor of mean CMC variability across unweighted networks; and most networks contain a subset of nodes with consistently high scores across nearly all centrality measures and another subset with consistently low scores.

Consistent with past findings [16,17], most CMCs were high, although there was considerable variability across networks (Fig. 1 and Fig S3). CMCs in weighted networks were only slightly weaker than their unweighted forms. Notably, the simplest and most popular measure of centrality, node degree, showed high correlations with most other centrality metrics, likely because a highly connected node is likely to be rated as central by other metrics (see discussion of random networks below). Degree may thus act as a useful first approximation of node centrality. Despite generally high CMCs, some measures showed low correlations with other metrics. For instance, Leverage and PageRank centrality were both highly correlated with each other but less so with other measures in both weighted and unweighted networks, possibly because these measures scale a node's importance in relation to the importance of its immediate neighbours, unlike other centrality measures. Bridging centrality and the participation coefficient also demonstrated weaker correlations with other measures.

We found density, global efficiency, modularity, majorization gap, and spectral gap were correlated with CMCs, which is in line with past findings [20,21,23,27]. Of these, the majorization has been most clearly linked to CMCs by theory [18–20]. However, our regression analysis revealed that the majorization gap was not a significant predictor of the unweighted mean within-network CMCs. The weak association between majorization gap and CMCs was confirmed by the analysis of surrogate data — where theory predicts that a lower gap should be associated with higher CMCs, our surrogates were characterized by CMCs despite having a comparable or larger gap relative to the observed networks. Recent work has also noted that the possible ranks each node could have on different centrality measure can be calculated from the neighbourhood inclusion preorder, and it is the preorder which the majorization gap is heuristically trying to assess. Indeed, our data suggests the majorization gap is likely not related to the exact centrality ranks each node will achieve, but rather the variability in possible ranks a node could have [19]. Our regression analysis also indicated that modularity was the only topological property to make a significant, unique contribution to mean CMC variation across networks. Networks with higher modularity than their matched surrogates also had weaker CMCs (and vice-versa). Modular networks provide greater opportunities to decouple local from global measures of centrality; they can also result in bottlenecks that can dissociate path-based from degree-based measures (e.g., Figure 1b). The net effect will be a reduction in mean CMCs.

We note that our empirical analysis measured global properties of network topology using methods that may only approximate the actual topology. For example the modularity of a network is highly dependent on the decomposition algorithm used [49], it is not clear how large the spectral gap needs to be for a network to be a good expander [21,23], and the majorization gap is a heuristic for quantify the distance of a network from a threshold graph, which itself is itself a heuristic to generate a network with perfect neighbourhood-inclusion preorder [20].

Hierarchical clustering of multivariate nodal centrality profiles indicated that two general clusters are present in nearly all networks: a subset of nodes scoring highly on nearly all centrality measures, representing a putative core, and a subset of nodes with low scores on nearly all measures, representing a putative periphery. Beyond these clusters, networks fell into one of two classes, such that they either shows a gradual progression moving from highly central core nodes to peripheral nodes, or a more clustered structure in which subsets of nodes had distinct centrality profiles. These intermediate clusters may define distinct nodes roles that cannot be identified through reliance on a single centrality measures. Networks with this structure tended to have higher modularity, or formed a ring with "tendrils" of nodes (i.e. Fig. S12). Together, these results suggest that multivariate centrality profiles may be particularly useful in characterizing nodes roles in networks with modular structure.

An unresolved question concerns the optimal set of centrality measures for such centrality profiling. We focused on a small subset of the >200 metrics that have been proposed, and a wider investigation of this issue is required. We note however, that a limitation of using hierarchical clustering to group nodes is that this approach is unlikely to place individual nodes (or small subsets of nodes) with a distinctive centrality profiles within a separate cluster. Indeed, we did find that some networks do contain a small number of nodes with highly discrepant scores across centrality measures (e.g., Fig. 7 and Figs. S12-15). Alternative clustering approaches may be better placed to delineate such nodes, which may play an important role in shaping network dynamics. Nonetheless, our basic approach demonstrates how a comparative approach to centrality analysis, as has been employed in other domains [30], can yield useful insights into the roles of different nodes within a network.

**Acknowledgements**

B.F. was supported by a National Health and Medical Research Council Early Career Fellowship (ID: 1089718); AF was supported by the Australian Research Council (ID: FT130100589) and National Health and Medical Research Council (IDs: 1146292, 1050504, 1104580 and 1066779). The authors also wish to thank Ernesto Estrada for useful feedback on the manuscript.

**Author Contributions Statement**

S.O, B.F and A.F developed the concept of the study and implemented the design. S.O analysed the data and produced all figures. S.O, A.A, L.P and A.F constructed the methodology needed to create the brain networks. S.O and C.S developed data management and analysis strategies. S.O, B.F and A.F wrote the main manuscript text. All authors reviewed the manuscript.

**Competing interests**

The authors declare no competing interests.

**Title: Consistency and differences between centrality measures across distinct classes of networks**

Authors: Stuart Oldham [1*], Ben Fulcher [1,2], Linden Parkes [1], Aurina Arnatkevičiūtė [1], Chao Suo [1], Alex Fornito [1]

[1] Brain and Mental Health Research Hub, School of Psychological Sciences and the Monash Institute of Cognitive and Clinical Neurosciences (MICCN), Monash University, Australia.

[2] School of Physics, University of Sydney, Australia.

Conflict of Interest: All authors report no conflict of interest.

**\*Corresponding author:**

Stuart Oldham

Brain and Mental Health Research Hub, School of Psychological Sciences and the Monash Institute of Cognitive and Clinical Neurosciences (MICCN), Monash University.

E: stuart.oldham@monash.edu

**Supplementary Methods**

**Centrality Definitions.** Each network is represented as an $N \times N$ adjacency matrix $A$ in which the element $A_{ij} = 1$ if nodes $i$ and $j$ are connected and $A_{ij} = 0$ otherwise. We denote the adjacency matrix of a weighted network $W$, where the element $W_{ij}$ encodes the weight of the edge between nodes $i$ and $j$. In the following, we present definitions of centrality measures for unweighted networks. Unless otherwise explicitly noted, these definitions were generalized to weighted networks simply substituting by $W_{ij}$ for $A_{ij}$.

**Degree/Strength (DC).** The simplest measure of centrality is degree centrality [1], defined as the number of edges attached to a node:

$$DC_i = d_i = \sum_{j \neq i} A_{ij}.$$

For weighted networks, we used the analogous measure of weighted degree, otherwise known as node strength $s$, which is the sum of all edge weights attached to a node,

$$DC_i = s_i = \sum_{j \neq i} W_{ij}.$$

**H-index Centrality (HC).** While commonly used to quantify the productivity and impact of a scientists' work, the h-index has also been recently applied as a centrality metric in complex network analysis [2]. If $\mathcal{N}_{\geq h}(i)$ is the set of neighbours of node $i$ that have a degree equal to or greater than $h$, the h-index of node $i$ can be defined as

$$HC_i = \max_{1 \leq h \leq d_i} \min(|\mathcal{N}_{\geq h}(i)|, h),$$

where $h$ is a value between one and the degree of node $i$. Thus, h-index of a node, $i$, is defined as the maximum value $h$ for which $h$ of node $i$'s neighbours have a degree of at least $h$. When weighted networks were used, $\mathcal{N}_{\geq h}(i)$ is the set of neighbours of node $i$ that have a strength equal to or greater than $h$.

**Leverage Centrality (LC).** Another centrality measure that considers the connections of a node's neighbours is leverage centrality [3]. Unlikely other centrality measures, leverage centrality can assign negative values to a node, indicating that node has less connections than its neighbours. In this case, a node is said to be influenced by its neighbours. Conversely a node with positive values has more connections than its neighbours, implying that it exerts influence overs its neighbours. Leverage centrality is defined as

$$LC_i = \frac{1}{d_i} \sum_{j \in \mathcal{N}(i)} \frac{d_i - d_j}{d_i + d_j},$$

where $\mathcal{N}(i)$ is the set of neighbours of node $i$. In weighted networks the equation becomes

$$LC_i = \frac{1}{d_i} \sum_{j \in \mathcal{N}(i)} \frac{s_i - s_j}{s_i + s_j}.$$

**Eigenvector Centrality (EC).** Eigenvector centrality assigns a high score to nodes that have high degree and/or have neighbours with high degree [4]. This measure is defined as the eigenvector, $v$, associated with the largest eigenvalue $\lambda_1$ of the adjacency matrix, and can be written as

$$EC_i = v_i = \frac{1}{\lambda_1} \sum_j A_{ji} v_j.$$

**Katz Centrality (KC).** In a connected network with a large, densely connected module, eigenvector centrality will assign high scores for nodes within the module and low (if not zero) scores for nodes outside the module; the measure thus becomes unsuitable for distinguishing nodes outside the module [5]. To overcome this, Katz centrality adds two parameters, $\alpha$ and $\beta$, to the definition of eigenvector centrality. The parameter $\alpha$ penalizes the contribution of distant dependencies (i.e. neighbours of neighbouring nodes) to a node's centrality score. The parameter $\beta$ assigns a specified amount of centrality to each node, thus ensuring every node as a non-zero centrality value [6]. As Katz centrality assigns every node a small amount of centrality, this ensures that highly connected nodes in other clusters are also assigned high centrality scores. Katz centrality can be written as

$$KC_i = \alpha \sum_j A_{ji} v_j + \beta,$$

or in matrix form as

$$KC = \vec{\beta}(I - \alpha A)^{-1},$$

where $\vec{\beta}$ is a vector of size $N$ with each element equal to $\beta$ and $I$ is the identity matrix of $A$. For all analyses, $\alpha$ was set to be 10% less than the inverse of the largest eigenvalue (as typically a value close to the largest eigenvalue is used) [7] of the network and $\beta$ was set to 1.

**PageRank centrality (PR).** With Eigenvector and Katz centrality, low degree nodes may receive a high score simply because they are connected to very high degree nodes, despite having low degree. PageRank centrality corrects for this behaviour by scaling the contribution of node $i's$ neighbours, $j$, to the centrality of node $i$ by the degree of $i$ [8],

$$PR_i = \alpha \sum_j A_{ji} \frac{v_j}{d_j} + \beta.$$

This definition can be written in matrix form as

$$PR = \vec{\beta}(I - \alpha D^{-1} A)^{-1},$$

where $D$ is a diagonal matrix and $D_{ii}$ is the degree of node $i$ (in weighted networks $S$ is used instead where the diagonal $S_{ii}$ is the strength of node $i$). Note the parameters $\alpha$ and $\beta$ have the same function as in Katz centrality. $\beta$ was set to 1 and $\alpha$ was set to 0.85 for all analysis.

**Closeness Centrality (CC).** Closeness centrality defines a node as central if it has a low average minimum path length to every other node in the network [9]. It is assumed that nodes with a short average path length to other nodes can spread or receive information in a relatively short amount of time. Since a smaller average path length indicates a more central node, the inverse is taken so that more central nodes are given higher values. This is defined by

$$CC_i = \frac{N}{\sum_j l_{ij}},$$

where $l_{ij}$ is the shortest topological distance between nodes $i$ and $j$. For weighted networks, $l_{ij}$ was computed using the weighted shortest path (the path with the smallest edge weight sum) using the inverse of the weight matrix (as larger weight indicates greater importance in all the weighted networks used here).

**Information centrality (IC).** This measure, also sometimes referred to as current-flow closeness centrality [10], considers all possible paths that could exist between two nodes and the overlap between these paths, and weights them per the amount of information that path contains[11]. The information in a path is defined as the inverse of the topological length of that path.

To estimate information centrality, we first define the matrix $C = (L + J)^{-1}$, where $L$ is the Laplacian of $A$ and $J$ is a $N \times N$ matrix with all elements equal to one. Information centrality is then defined as

$$IC_i = \left( C_{ii} + \frac{\sum_j C_{jj} - 2\sum_j C_{ij}}{N} \right)^{-1}.$$

In weighted networks $L$ is the Laplacian of $W$.

**Random-Walk Closeness Centrality (RWCC).** Random-walk closeness centrality (RWCC) measures the average amount of time it takes a random-walker starting at any node in the network to reach node $i$ [12,13] and is equal to the inverse of the average mean-first passage time (MFPT) to a specific node. The MFPT can be computed from the fundamental matrix $Z$

$$Z = (I - P + \Pi)^{-1},$$

where $I$ is the identity matrix, the transition matrix $P = D^{-1}A$ (or $P = S^{-1}W$ in weighted networks), and $\Pi$ is a $N \times N$ matrix where each column is the vector $\pi$ of steady state (also known as limiting) distribution probabilities of the transition matrix (such that $\Pi_{ij} = \pi_j$). The vector $\pi$ can be obtained by solving the system of linear equations $\pi P = \pi$ and $\sum_i^N \pi_i = 1$. The MFPT matrix $H$ can then be defined as

$$H_{ij} = \frac{Z_{jj} - Z_{ij}}{\pi_j}, i \neq j.$$

The element $H_{ij}$ is the MFPT from node $i$ to node $j$ [14,15]. RWCC is then simply calculated as

$$RWCC_i = \frac{N}{\sum_j H_{ji}}.$$

**Subgraph Centrality (SC).** Like other measures, subgraph centrality also counts the number of walks, but instead of counting walks to other nodes, this method considers closed walks (i.e. walk that begin and end at the same node) [16]. Thus, subgraph centrality measures how many subgraphs, defined by closed walks, that node belongs to, with smaller subgraphs being assigned more importance. Longer walks (and hence larger subgraphs) are penalized by weighting each walk by factor $\frac{1}{n!}$, where $n$ is the length of the walk. Thus, subgraph centrality can be computed as

$$SC_i = \sum_{n=0}^{\infty} \frac{[A^n]_{ii}}{n!} = [e^A]_{ii}.$$

In weighted networks, the reduced adjacency matrix $S^{-\frac{1}{2}}WS^{-\frac{1}{2}}$ is used instead of $A$ [17].

**Total Communicability centrality (TCC).** Another measure which accounts for all possible walks is total communicability [18]. It is like subgraph centrality in that it considers all possible walks and walks are weighted by the inverse of the factorial of their length, but instead of just accounting for closed walks, this method also considers walks to other nodes. It is similar to information centrality in that it considers indirect routes of communication between nodes, but with walks instead of paths.

The total communicability of node $i$ can be expressed as the sum of all weighted walks to all other nodes

$$TCC_i = \sum_{n=0}^{\infty} \sum_{j} \frac{[A^n]_{ji}}{n!} = \sum_{j} [e^A]_{ji}.$$

As with subgraph centrality, the reduced adjacency matrix $D^{-\frac{1}{2}}wD^{-\frac{1}{2}}$ is used instead of $A$ when the network is weighted.

**Laplacian centrality (LAPC).** A node can be thought of as topologically central if its removal would impair the network in some manner. One way this can be quantified is to measure the Laplacian energy of a network

$$LAPC_i = 4NW_2^C(i) + 2NW_2^E(i) + 2NW_2^M(i),$$

where $NW_2^C(i) = \sum_{j \in \mathcal{N}(i)} W_{ij}^2$ which is the number of closed 2-walks involving node $i$, $NW_2^E(i) = \sum_{j \in \mathcal{N}(i)} \left( \sum_{k \in \mathcal{N}(j), i \neq k} W_{ij} W_{kj} \right)$ which is the number of non-closed 2-walks where $i$ is one of the end nodes, and $NW_2^M(i) = \frac{1}{2} \sum_{j,k \in \mathcal{N}(i)} W_{ij} W_{ik}, j \neq k$ which is the number of non-closed 2-walks containing node $i$ as the middle node [19]. This definition is applicable weighted networks; in unweighted networks [20], the equation simplifies to

$$LAPC_i = d_i^2 + d_i + 2 \sum_{j \in \mathcal{N}(i)} d_j.$$

**Shortest-path Betweenness centrality (BC).** Shortest-path Betweenness centrality defines nodes as central if they lie on many shortest-paths between other pairs of nodes [21]. The measure assumes that nodes with high betweenness act as putative information-processing bottlenecks in the network.

If $g_{jk}$ is the number of shortest paths (or geodesic paths) between nodes $p$ and $q$ and $g_{pq}(i)$ is the number of shortest paths between nodes $p$ and $q$ which pass through node $i$, then the betweenness centrality of node $i$ is

$$BC_i = \sum_{p \neq i, p \neq q, q \neq i} \frac{g_{pq}(i)}{g_{pq}}.$$

In weighted networks the shortest path is calculated as the path with the smallest edge weight sum. As all weighted networks used here define a larger edge weight as being of more importance, the adjacency matrix was inverted prior to the shortest paths being calculated.

**Random-walk betweenness centrality (RWBC).** The classical definition of betweenness considers only the shortest path between two nodes. However, nodes can interact through alternative routes. One way to assess such routes is to use the movements of a random-walker on a network and count how many times it passes through a given node as it travels between two others [22].

The movement of a random-walker is comparable to an electric current flowing through the network (which leads to random-walk betweenness sometimes being referred to as current flow betweenness) where each edge is a resistor and each pair of nodes is acting as a source and drain. Thus, random-walk betweenness centrality can be calculated as the average current flowing through node $i$ over all pairs of node sources $p$ and drains $q$. The currents are calculated by

$$I_i^{(pq)} = \begin{cases} 1, & i = p, q \\ \frac{1}{2}\sum_j A_{ij}|T_{ip} - T_{iq} - T_{jp} + T_{jq}|, & i \neq p, q \end{cases}.$$

$T$ is calculated by first removing row and column $v$ from the matrix $D$ and $A$, giving $D_v$ and $A_v$ respectively, and calculating the matrix $(D_v - A_v)^{-1}$. Column and row $v$ are added back into this matrix with values all equal to zero to produce the matrix $T$. The random-walk betweenness centrality of node $i$ is then calculated as

$$RWBC_i = \frac{\sum_{p<q} I_i^{(pq)}}{\frac{1}{2}N(N-1)}.$$

In weighted networks $D$ and $A$ are substituted for $S$ and $W$ respectively. Note that this measure only considers the net-forward movement of a random-walker

**Communicability betweenness centrality (CBC).** As previously mentioned, another way of assessing alternative routes between nodes is communicability. Communicability betweenness considers the number of walks between every pair of nodes in which a given node participates [23].

The communicability betweenness of node $i$ is

$$CBC_i = \frac{1}{\acute{C}}\sum_p \sum_q \frac{G_{piq}}{G_{pq}}, p \neq q, q \neq i,$$

where $G_{piq} = (e^A)_{pq} - (e^{A+A'(i)})_{pq}$ is the number of walks between nodes $p$ and $q$ involving node $i$, with $A'(i)$ is the adjacency matrix with all rows and columns apart from $i$ being zero, $G_{pq} = (e^A)_{pq}$ is the number of closed walks starting at $p$ and ending at $q$, and $\acute{C} = (N-1)^2 - (N-1)$ is a normalising term. As with subgraph centrality, the reduced adjacency matrix is used instead of $A$ for weighted networks.

**Bridging centrality (BridC).** Noting that many different centrality measures are heavily influenced by nodal degree, Hwang and colleagues devised a measure known as bridging centrality, which aims to identify nodes that are central because they connect different communities/modules [24]. The measure is obtained by scaling a node's shortest-path betweenness centrality by the nodes bridging coefficient, which is defined as

$$Bc_i = \frac{d_i^{-1}}{\sum_{j \in \mathcal{N}(i)} d_j^{-1}}.$$

The coefficient quantifies the extent to which a node's neighbours have a higher degree. Bridging centrality quantifies how many paths between highly connected nodes pass through a given node,

$$BridC_i = BC_i \times Bc_i.$$

**Participation coefficient (PC).** Most real-world networks are modular [25] and nodes that play an important integrative role in the network will connect to a diverse range of modules. This distribution of a node's connections across modules can be quantified using the participation coefficient [26]

$$PC_i = 1 - \sum_{m=1}^{M} \left(\frac{d_i(m)}{d(i)}\right)^2,$$

where $M$ is the number of modules in the network and $d_i(m)$ is the number of connections of node $i$ that links to nodes in module $m$. The module membership of each node was determined via consensus clustering (see below).

**Global network properties.**

**Assortativity.** When nodes tend to be connected with other similar nodes, this property is known as assortativity. Commonly, this is defined in terms of node degree such that a network that is highly assortative has nodes of a similar degree connected to each other [7,27,28]. This is defined as the Pearson correlation between the degree of connected nodes and is calculated by

$$r = \frac{E^{-1}\sum_{i,j} d_i d_j - \left[E^{-1}\sum_{i,j} \frac{d_i + d_j}{2}\right]^2}{E^{-1}\sum_{i,j} \frac{d_i^2 + d_j^2}{2} - \left[E^{-1}\sum_{i,j} \frac{d_i + d_j}{2}\right]^2},$$

where $d$ can be substituted with $s$ in weighted networks.

**Clustering.** Clustering was defined as the average number of pairs of neighbours of a node that are connected [29]

$$Cl = \frac{1}{N}\sum_i \frac{2t_i}{d_i(d_i - 1)},$$

where $t_i$ is the number of closed triangles attached to $i$. In weighted networks the clustering coefficient represents how equal the weights in the closed triangle are to the maximum edge weight in the network [30]

$$Cl^w = \frac{1}{N}\sum_i \frac{2}{d_i(d_i - 1)} \sum_{j,k} (\widehat{W}_{ij}\widehat{W}_{kj}\widehat{W}_{ik})^{\frac{1}{3}},$$

where $\widehat{W}_{ij}$ is where the edge weight of $i$ and $j$ have been scaled to the maximum edge weight in $W$.

**Network density.** Density is the proportion of all possible connections in a network that exist. It is defined for undirected networks as

$$\kappa = \frac{2E}{N(N-1)},$$

where $E$ is the number of edges in the network and $N$ is the number of nodes.

**Efficiency.** The average path length $L$ of a network indicates the average length of the shortest-paths between nodes. As a lower average path length indicates that less traversals have to be made to move from one node to another, in which case a network is considered to be more efficient. This is simply termed global efficiency [31], which is simply the reciprocal of $L$.

**Diffusion efficiency.** While efficiency quantifies the length of the average shortest-path, diffusion efficiency quantifies the average length of random-walks between nodes [14]

$$E_{diff} = \frac{1}{N(N-1)} \sum_{i \neq j} \frac{1}{H_{ij}}.$$

**Majorization gap.** The majorization gap quantifies the distance between an empirical network and an idealized network, called a threshold graph, in which all centrality measures rank nodes in the same way [32]. In their analysis of social networks, Schoch and Brandes [33] argued that a given metric should only be considered a measure of centrality if it preserves a property called the neighbourhood-inclusion preorder. If the neighbours of node $j$ are a subset of the neighbours of node $i$, then node $i$ is said to dominate node $j$, and must have a greater or equivalent level of centrality. The neighbourhood inclusion preorder is the rank ordering of nodes in terms of these dominance relationships, such that nodes that are not dominated by any others are ranked first, and are thus more central, and nodes that are dominated by many others are ranked last (and are thus least central; e.g., Fig. S1). This preorder is complete – each node is either dominated by another or not – in a class of networks known as a threshold graphs. Such graphs are formed by adding nodes to a network, one at a time, such that the new node either connects to all existing nodes or connects to no other nodes (see Fig. S2 for an example). In these networks, all centrality measures rank nodes in the same order, and this order is perfectly concordant with the neighbourhood-inclusion preorder.

Schoch and Brandes [33] argue that higher CMCs will be apparent in an empirical network if it is topologically similar to a comparable threshold graph. This similarity can be quantified using the *majorization gap*, which estimates the number of edges that must be rewired to transform a network into a threshold graph [32]. The majorization gap is defined as

$$Mgap = \frac{1}{2} \sum_{k=1}^{n} \max\{d'_k - d_k, 0\},$$

where $d$ is the degree sequence defined by

$$d = [d_1, d_2, \ldots, d_n] \text{ with } d_1 \geq d_2 \geq \ldots \geq d_n,$$

and $d'$ is the corrected conjugated sequence. For a node in position $k$ in the degree sequence, the conjugated sequence describes how many nodes before it in the degree sequence have a degree greater than $k - 1$, and how many nodes following it in the degree sequence have a degree greater than $k$; formally,

$$d'_k = |\{i: i < k \land d_i \geq k-1\}| + |\{i: i > k \land d_i \geq k\}| \text{ with } 1 \leq k \leq n.$$

To facilitate comparison across networks, $Mgap$ was normalised by the number of edges in the network [32].

**Modularity.** Modularity was quantified using the widely-used $Q$ metric, first proposed by Newman and Girvan [34] and defined as:

$$Q = \frac{1}{2E}\sum_{i,j}(A_{ij} - e_{ij})\delta(m_i, m_j),$$

where $E$ is the number of edges in the network, $e_{ij} = \frac{k_i k_j}{2E}$ and $\delta(m_i, m_j)$ is the Kronecker delta function which is equal to one if nodes $i$ and $j$ are part of the same module and zero otherwise. For weighted networks, $E$ was replaced with the total weight of unique edges in the network and $k$ was replaced with the sum of all edge weights attached to a node [35]. When $Q > 0$, the network exhibits greater connectivity within modules than expected by chance, under the configuration model. Modules were identified using the Louvain algorithm [36] combined with a consensus clustering procedure to address algorithmic degeneracy [37]. The Louvain algorithm was run 50 times on each network's adjacency matrix, giving 50 partitions of the network along with an associated modularity quality score $Q$. A consensus classification matrix was computed in which each element indicated the fraction of times two nodes in the network had been assigned to the same module over the 50 iterations. This matrix was weighted by $Q$, such that higher quality partitions received higher weighting than lower quality partitions. A threshold of 0.4 (a value in the range 0.3-0.7 is recommend for use in Louvain clustering) [38] was applied to the matrix so that values below this threshold were set to 0. Louvain community detection was run on the thresholded matrix 50 times to produce another set of 50 partitions (approximately 50 iterations are required to produce an optimal partition). This process was repeated until the final consensus matrix resulted in node pairs either always being assigned to the same module or never being assigned to the same module[38].

**Spectral gap**. The absolute difference between the principal and second largest eigenvalues of $A$ is the spectral gap, i.e. $|\lambda_1 - \lambda_2|$ where $\lambda_1 \geq \lambda_2 \geq \cdots \geq \lambda_N$.. A large spectral gap is suggestive of a network having good expander properties, whereby it is both sparsely yet well connected [18,39,40]. Such networks are shown to have high correlations between walk-based centrality measures. In this paper we compute the spectral gap as the ratio $1 - \frac{\lambda_2}{\lambda_1}$ (so that a larger value indicates a larger spectral gap) to provide a normalized value to allow comparison across networks [32].

**Human Brain Network Construction.**

Human structural brain networks were created from the Human Connectome Project (HCP) [41]. The HCP dataset comprised diffusion-weighted MRI (1.25 mm³ voxel size, TE/TR = 89.5/5520ms, FOV = 210 × 180 mm, 90 directions with b = 1000, 2000, 3000 s/mm², six b = 0) and T1-weighted MRI (0.7 mm3 voxel size, TR/TE = 2400/2.14ms, FOV of 224x224 mm) for 100 unrelated participants (54 males, 46 females, age range of 22-35 years) from the 500 data release. The data were acquired using a customized head coil (100 mT/m maximum gradient strength and a 32 channel head coil) on a 3T scanner located at Washington University, St Louis. All HCP data had previously gone through an extensive pre-processing pipeline [42]. Pre-processing for structural images included bias field correction, registration from native to MNI space, and segmentation of the volume into 34 cortical and seven subcortical regions for each hemisphere to produce the same 82 node parcellation as mentioned above. The HCP Diffusion MRI processing pipeline included normalization of b0 image intensity across runs, and correction for EPI susceptibility, eddy-current-induced distortions, slice dropouts, gradient-nonlinearities and subject motion.

Network nodes were defined using a recently-developed, data-driven parcellation of the cortex into 360 regions (180 per hemisphere) [43], This cortical parcellation was combined with a segmentation of seven thalamic [44,45] and three striatal [46] regions to produce a whole-brain parcellation of 380 nodes.

Diffusion images were processed using MRtrix3 [47] and the FMRIB Software Library[48]. From the corrected diffusion data, the major eigenvectors of the diffusion tensor and fibre orientation distributions (FODs) were extracted and used to conduct tractography with the Fibre Assignment by Continuous Tracking (FACT) algorithm. FACT propagates streamlines that track the trajectory of white matter tracts by following the primary direction of water diffusion at each voxel [49,50]. A total of 10 million streamlines were generated. Anatomically Constrained Tractography was employed alongside FACT using the tissue-segmented T1-weighted image to ensure that the generated streamlines were biologically accurate [51]. Dynamic seeding, where streamlines are sampled on a probabilistic basis of the relative difference between the estimated fibre density (calculated from the FOD) and current streamline reconstruction, was also employed when generating streamlines to ensure adequate sampling from across the entire brain [52]. Whole-brain tractograms were then re-weighted using Spherically Informed Filtering of Tractograms 2 (SIFT2) [52]. This algorithm adjusts streamline weights so that they more accurately represent the underlying diffusion signal, and thus provide a more physiologically meaningful measure of inter-regional connectivity.

In each subject, the parcellation and tractogram were combined to produce a network map of white matter tract connectivity. The start and end points of streamlines were assigned to the closest region within a 5mm radius. This was performed for each of the 100 participants. A single group-average connectome was then created using a consistency threshold. Specifically, for each edge, we estimated the coefficient or variation across participants and retained the 5.23% most consistent edges [53]. This consistency threshold was selected based on the average density observed across individuals in the dataset.

**Surrogate Networks.**

**Unconstrained surrogates.** For the unconstrained surrogate networks, we seek graphs that match only the size and density of each real-world network. A typical random network of $N$ nodes is generated by randomly allocating a set number of edges $E$ between pairs of nodes, or by forming edges between pairs of nodes with a given probability. However, such random networks are likely to be disconnected when the density is less than $\frac{\ln N}{N}$ [54]. Several networks had a density below this threshold. To ensure generation of a connected unconstrained/random network, we start by generating a random minimum spanning tree (MST) and then add edges at random. The procedure is as follows for a network of $N$ nodes and $E$ edges:

1. Each of the $N$ nodes are listed as undiscovered and no edges are placed in the network.
2. A node is chosen at random for a random-walker to start. The random-walker can move from one node to any other (except to the node it is currently positioned on). This first node that was chosen is labelled as being discovered.
3. When the random-walk reaches an undiscovered node, that node is now labelled as discovered. An edge is added to the network connecting the previous node the random-walk was on and the current, newly discovered node.
4. The random-walk continues until all nodes are discovered. The $N-1$ edges that have been added to the network form the MST.
5. From all remaining non-existent edges, a total of $E-N-1$ are selected uniformly at random and added in, leaving a connected random network.

This method of generating the MST ensures that the tree is generated uniformly at random[55]. Thus, when other edges are added at random, this network will have all the expected properties of a random network generated with standard algorithms. To generate weighted surrogates, edge weights of the original network were randomly assigned to edge in the connected surrogate graph.

**Constrained surrogates.** Constrained surrogates were generated using functions in the Brain Connectivity Toolbox in MATLAB [56]. The Maslov-Sneppen algorithm [57] was used for the unweighted networks to create a surrogate that preserves the number of nodes, number of edges, and degree distribution of the original network, without any fragmentation. For weighted networks, an algorithm that additionally preserves (approximately) the strength distribution of the original network was used [58].

| Variable | β | SE | $R^2$ | dfE |
|---|---|---|---|---|
| Model 1 | 0.869** | 0.104 | 0.505 | 205 |
|   Assortativity | -0.054 | 0.028 | 0.133 | |
|   Clustering | -0.033 | 0.036 | 0.063 | |
|   Efficiency | 0.192 | 0.117 | 0.114 | |
|   Majorization gap | -0.062 | 0.067 | 0.065 | |
|   Modularity | -0.308** | 0.092 | 0.228 | |
|   Spectral gap | -0.041 | 0.055 | 0.052 | |
| Model 2 | 0.608** | 0.070 | 0.517 | 206 |
|   Assortativity | -0.039 | 0.029 | 0.095 | |
|   Clustering | -0.106** | 0.029 | 0.244 | |
|   Efficiency | 0.488** | 0.079 | 0.395 | |
|   Majorization gap | -0.095 | 0.067 | 0.097 | |
|   Spectral gap | 0.057 | 0.047 | 0.084 | |
| Model 3 | 0.816** | 0.087 | 0.546 | 206 |
|   Assortativity | -0.068** | 0.024 | 0.198 | |
|   Clustering | -0.022 | 0.034 | 0.044 | |
|   Efficiency | 0.223* | 0.112 | 0.137 | |
|   Modularity | -0.320** | 0.091 | 0.239 | |
|   Spectral gap | -0.023 | 0.051 | 0.031 | |

**Table S1. General Linear Model of network properties predicting unweighted mean-within CMCs.** Density and diffusion efficiency were not included as they displayed non-linear relationships. * $p < .05$. ** $p < .01$.

| Variable | β | SE | $R^2$ | dfE |
|---|---|---|---|---|
| Model 1 | 0.867** | 0.098 | 0.844 | 33 |
|   Assortativity | -0.215** | 0.069 | 0.484 | |
|   Global efficiency | -0.171* | 0.073 | 0.381 | |
|   Majorization gap | -0.241 | 0.121 | 0.332 | |
|   Modularity | -0.135 | 0.155 | 0.153 | |
|   Spectral gap | 0.033 | 0.110 | 0.054 | |
| Model 2 | 0.814** | 0.077 | 0.880 | 34 |
|   Assortativity | -0.213** | 0.068 | 0.477 | |
|   Efficiency | -0.169* | 0.073 | 0.373 | |
|   Majorization gap | -0.317** | 0.083 | 0.554 | |
|   Spectral gap | 0.087 | 0.091 | 0.165 | |
| Model 3 | 0.854** | 0.101 | 0.826 | 34 |
|   Assortativity | -0.266** | 0.066 | 0.573 | |
|   Efficiency | -0.156* | 0.076 | 0.336 | |
|   Modularity | -0.359** | 0.111 | 0.489 | |
|   Spectral gap | 0.038 | 0.115 | 0.057 | |

**Table S2. General Linear Model of network properties predicting weighted mean-within CMCs.** Density, diffusion efficiency, and clustering were not included as they displayed non-linear relationships. * $p < .05$. ** $p < .01$.

**Supplementary references**

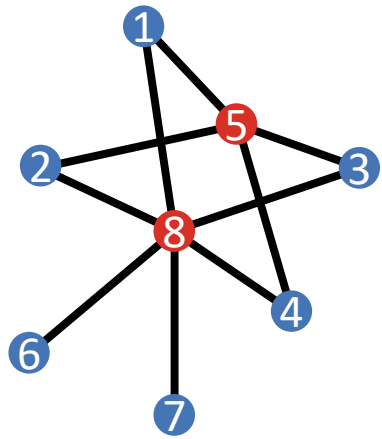

**Figure S1. A threshold graph.** A threshold graph is formed by adding in nodes one at a time in one of two ways: a node can either be added in forming no connections (blue nodes) or a node can be added forming connections to all existing nodes (red nodes). The number in each node is the order in which it was added into the network.

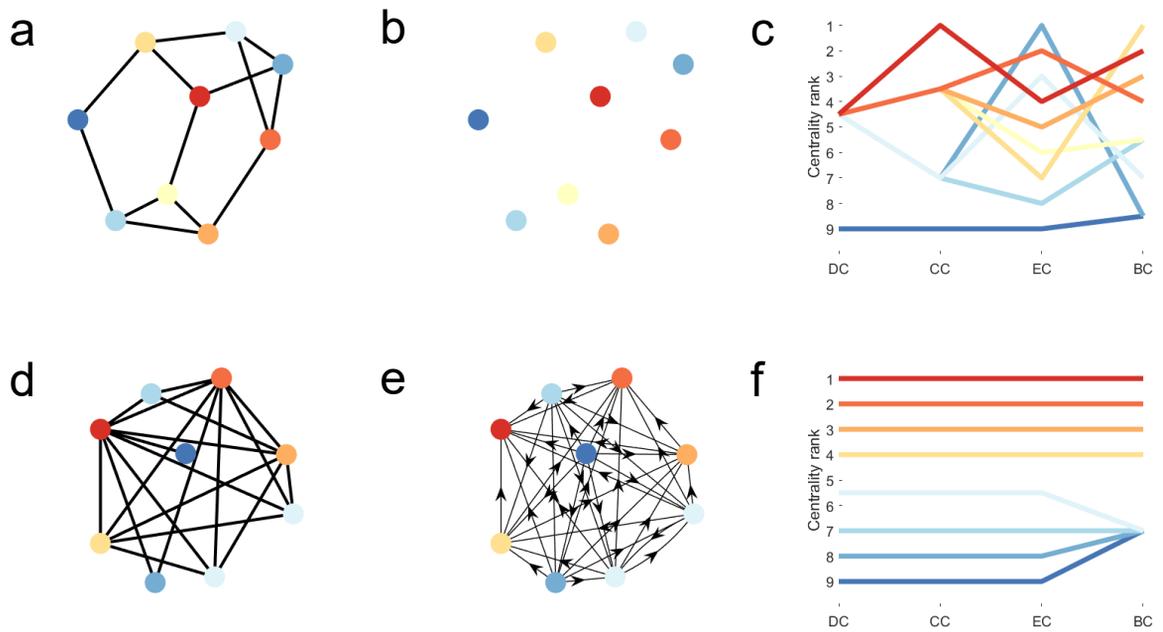

**Figure S2. The neighbourhood-inclusion pre-order and centrality ranks.** (a) and (d) shows a network which demonstrates the neighbourhood-inclusion pre-order and (b) and (e) shows the dominance relation between nodes (a directed edge indicates that the source node is dominated by the target, that is all the neighbours of the source node are a subset of the neighbours of the target node) for the respective network. The network in (a) has no dominance relationship thus has an inconsistent ranking of nodes by four different centrality measures: degree (DC), closeness (CC), eigenvector (EC), and betweenness (BC), as shown in (c). Conversely the network in (d) has a complete neighbourhood-inclusion pre-order (each node either dominates or is dominated any other node in the network) and thus all centrality measures will give the same ranking to a node, as shown in (f). Note that the network in (d) is a threshold graph. A similar figure is presented in Schoch and Brandes (33).

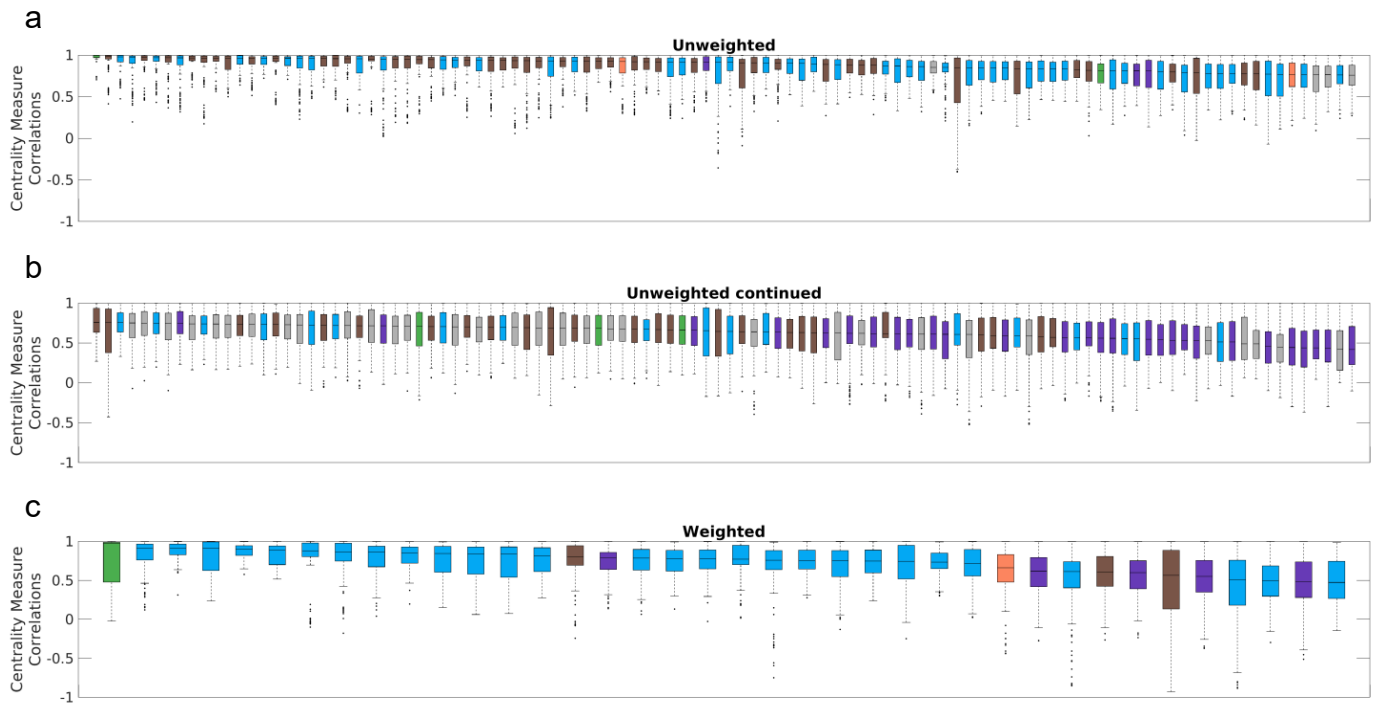

**Figure S3. Distributions of Centrality Measure Correlations (CMCs) in each network.** The distribution of Spearman correlation coefficient between every pair of centrality measures, CMCs, in each network is represented as a boxplot. Distributions for unweighted networks are split across (a) and (b), while distributions for weighted networks are shown in (c). Networks are coloured by their natural category (blue = social, grey = technological, brown = biological, orange = informational, purple = transportation; green = economic). Networks have been ordered from highest (left) to lowest (right) median CMC.

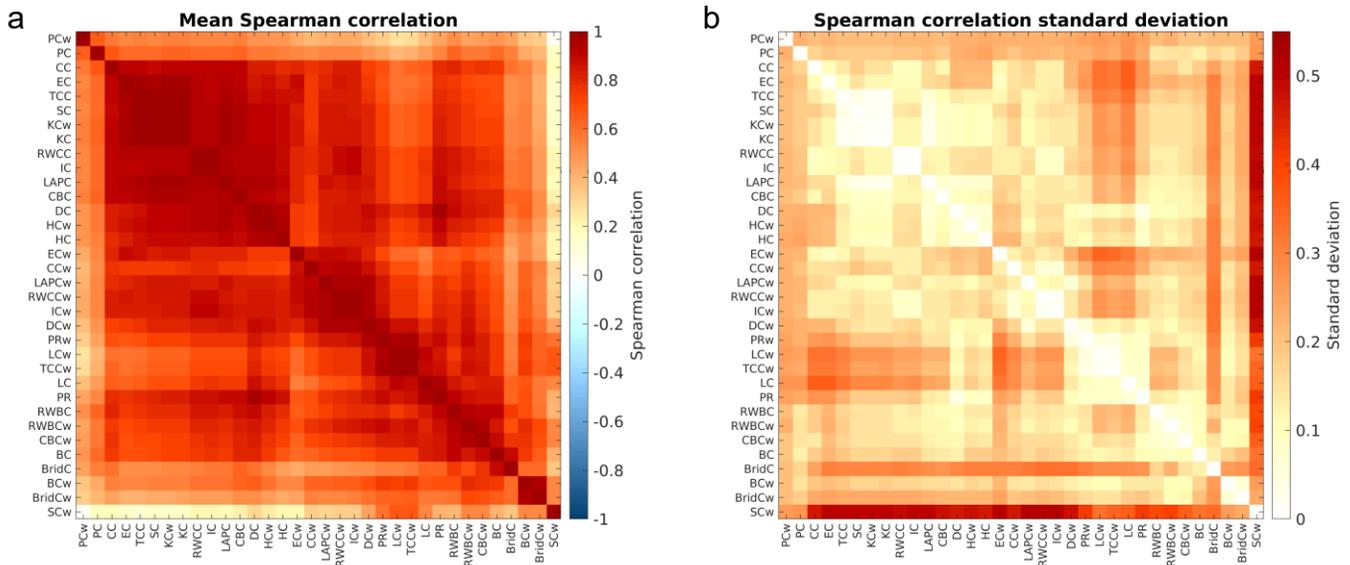

**Figure S4. Mean and standard deviation of between-network CMCs for unweighted and weighted measures.** This figure shows the between-network CMC mean (a) and standard deviation (b) for unweighted and weighted measures, as calculated on the 40 weighted networks. Both weighted and unweighted measures were generally highly correlated with each other, and unweighted measures were more highly intercorrelated than weighted measures. DC = Degree centrality; EC = Eigenvector centrality; KC = Katz centrality; PR = PageRank centrality; LC = Leverage Centrality; HC = H-index centrality; CC = Shortest-path closeness centrality; SC = Subgraph centrality; PC = Participation coefficient; TCC = Total communicability centrality; RWCC = Random-walk closeness centrality; BC = Shortest-path betweenness centrality; CBC = Communicability betweenness centrality; RWBC = Random-walk betweenness centrality; LAPC = Laplacian centrality; BridC = Bridging centrality. A "w" next to the abbreviated name for the centrality measure indicates it is the weighted version.

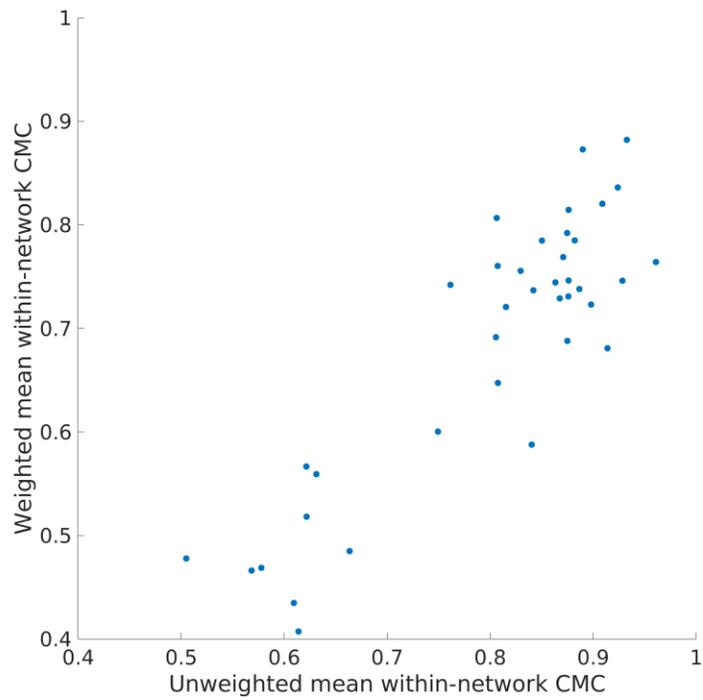

**Figure S5. Scatter plot of the mean within-network CMCs for unweighted and weighted measures**. For the 40 networks with edge weights the mean within-network CMC was calculated separately for unweighted and weighted measures, and then these values were plotted against each other. There is a strong relationship between the two indicating higher correlations between unweighted measures is mirrored by higher correlations in weighted measures.

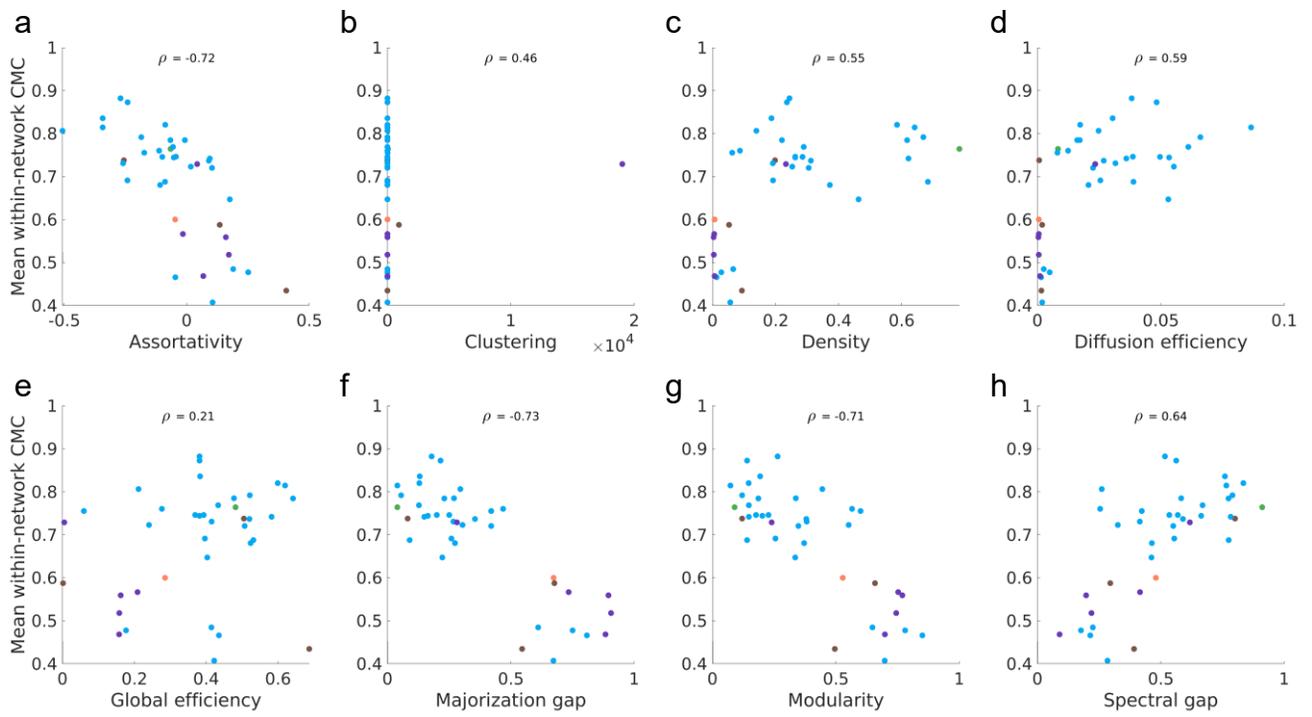

**Figure S6. Association between mean within-network CMC and network properties in weighted networks.** The association between the mean within-network CMC (the average of all CMCs within a single network) and each of the global topological properties. Networks are coloured by their natural category (blue = social, grey = technological, brown = biological, orange = informational, purple = transportation; green = economic)

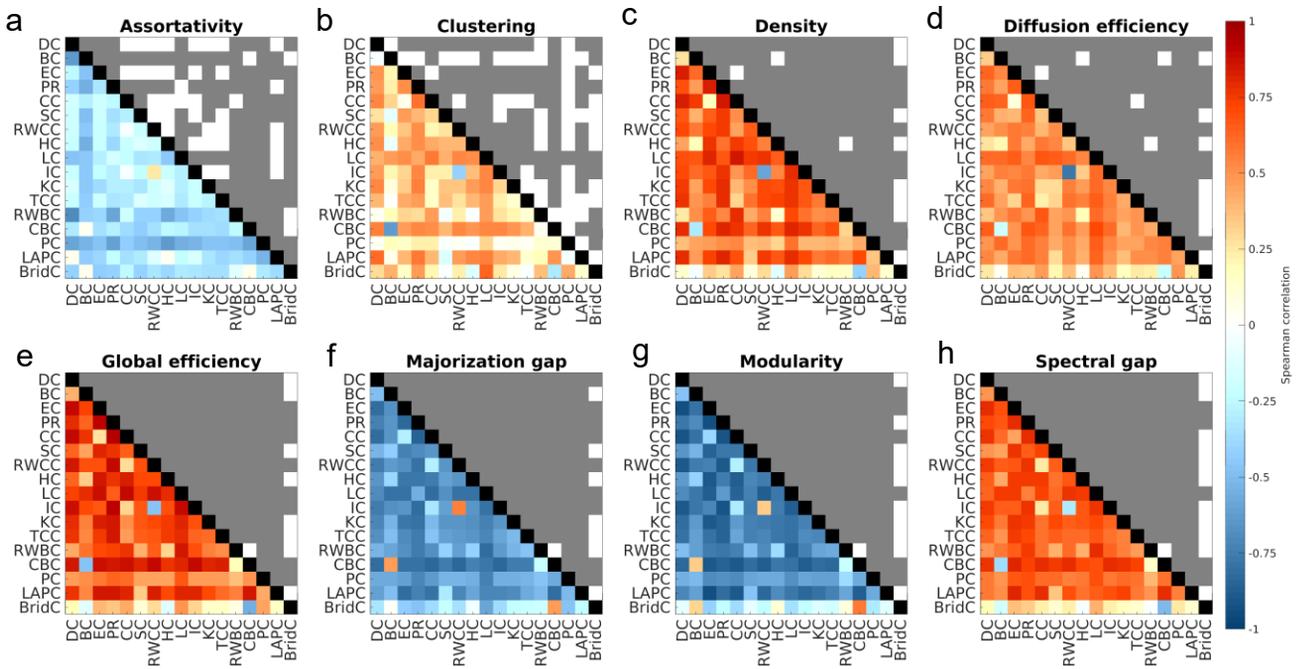

**Figure S7. Association between each CMC and global topology in unweighted networks.** The lower matrix indicates the value of the Spearman correlation between a CMC and a network property. The upper matrix indicates if this correlation was significant (grey) or not (white) when Bonferroni corrected for 136 combinations of centrality measures. This result shows the strength of individual CMCs was correlated with specific network properties.

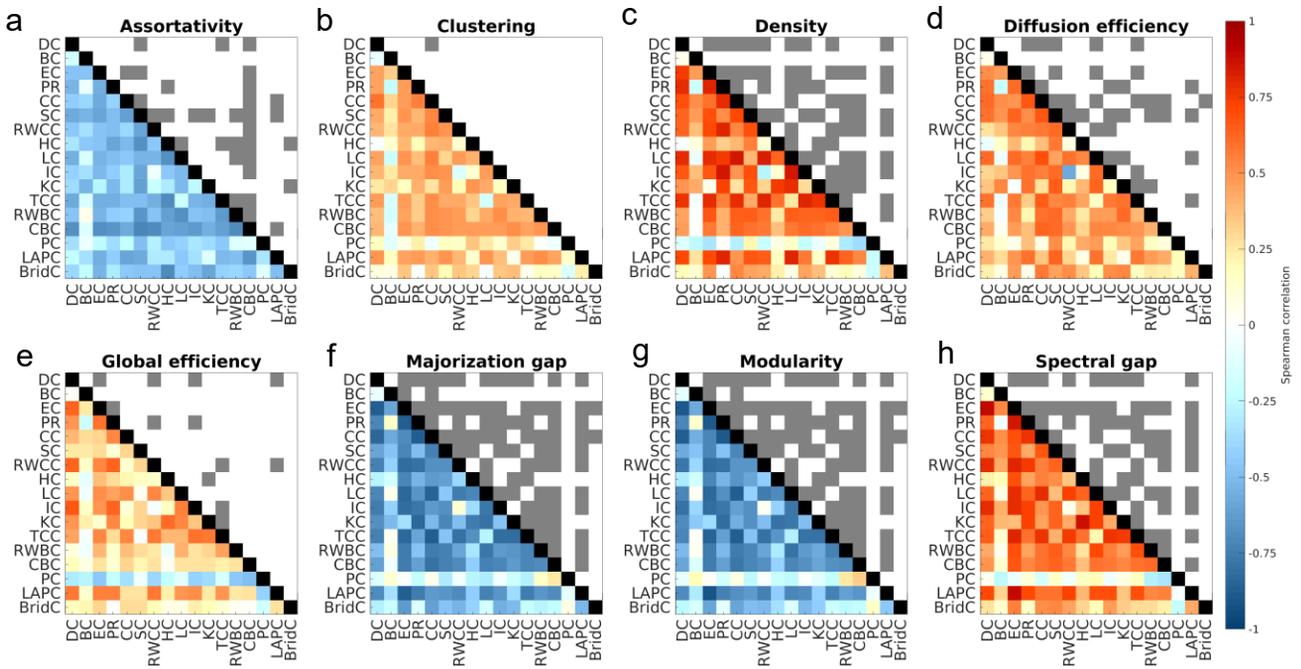

**Figure S8. Association between each CMC and global topology in weighted networks.** The lower matrix indicates the value of the Spearman correlation between a CMC and a network property. The upper matrix indicates if this correlation was significant (grey) or not (white) when Bonferroni corrected for 136 combinations of centrality measures. This result shows the strength of individual CMCs was correlated with specific network properties.

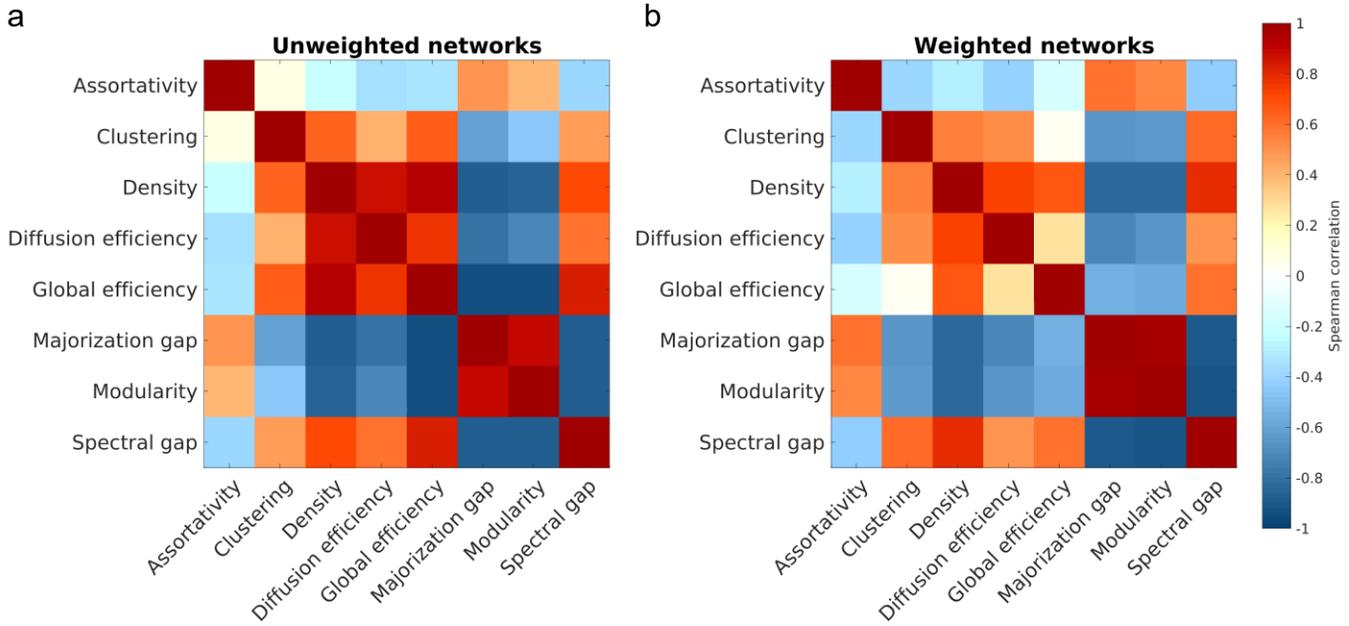

**Figure S9. Correlations between network properties in unweighted and weighted networks.** (a) shows the Spearman correlations between each network property in the unweighted networks, while (b) shows the Spearman correlations between each network property in the weighted networks.

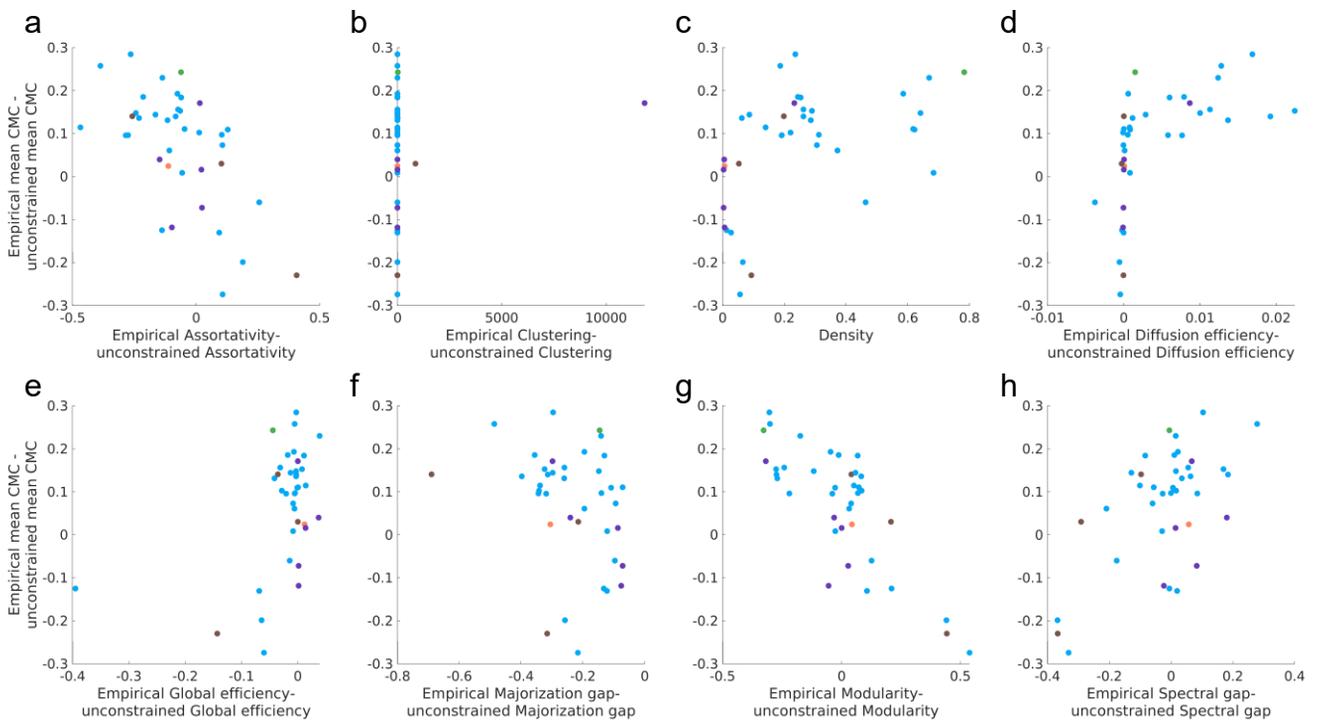

**Figure S10. Difference between weighted empirical and unconstrained surrogates in mean within-network CMC and network properties.** The y-axis of each plot shows the difference between the empirical networks and unconstrained surrogates mean within-network CMC. The x-axis shows the difference between the empirical networks and unconstrained surrogates on a particular property (except for (c) as the unconstrained surrogates have the same density as the empirical network). On both axis, except for the x-axis in (c), a negative value indicates the empirical network had a lower value than the mean value of the surrogates, while a positive value indicates the empirical networks had a larger value. Points are colored by the natural category of the empirical network (blue = social, grey = technological, brown = biological, orange = informational, purple = transportation; green = economic).

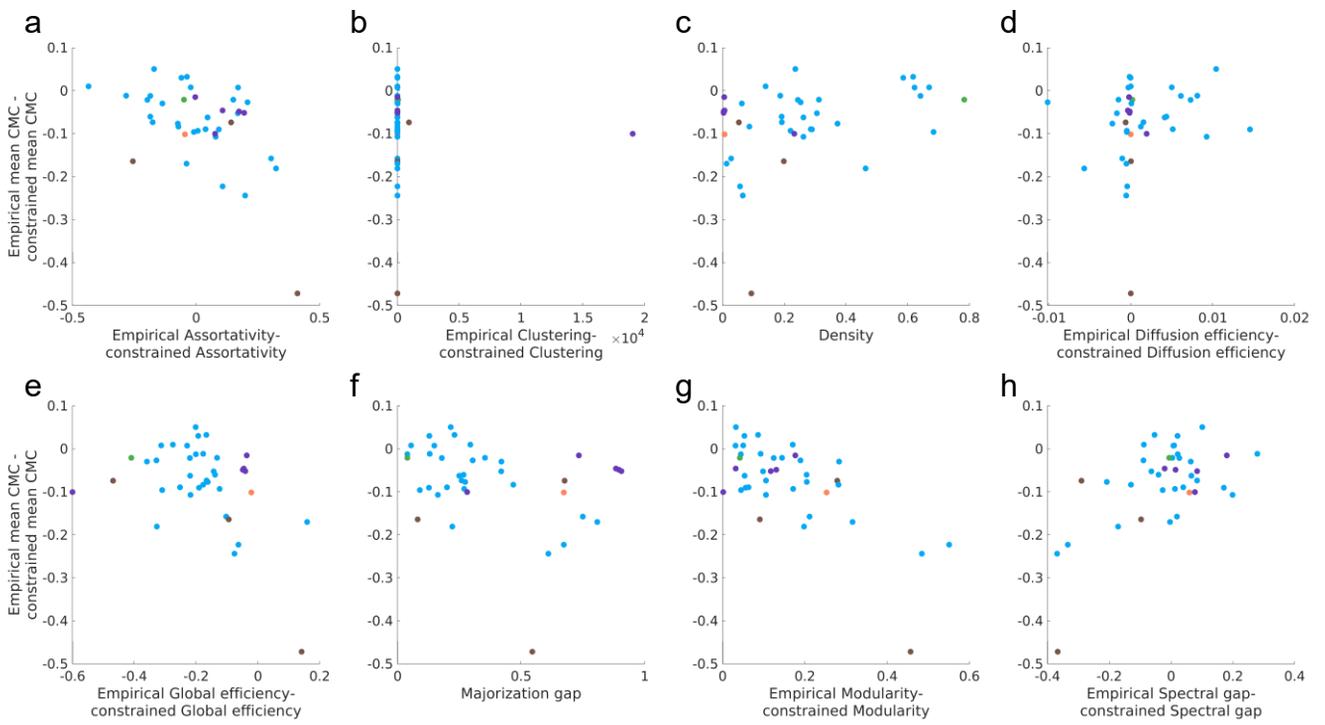

**Figure S11. Difference between weighted empirical and constrained surrogates in mean within-network CMC and network properties.** The y-axis of each plot shows the difference between the empirical networks and constrained surrogates mean within-network CMC. The x-axis shows the difference between the empirical networks and constrained surrogates on a particular property (except for (c) and (f) as the constrained surrogates have the same density and majorization gap as the empirical network). On both axis, except for the x-axis in (c) and (f), a negative value indicates the empirical network had a lower value than the mean value of the surrogates, while a positive value indicates the empirical networks had a larger value. Points are colored by the natural category of the empirical network (blue = social, grey = technological, brown = biological, orange = informational, purple = transportation; green = economic).

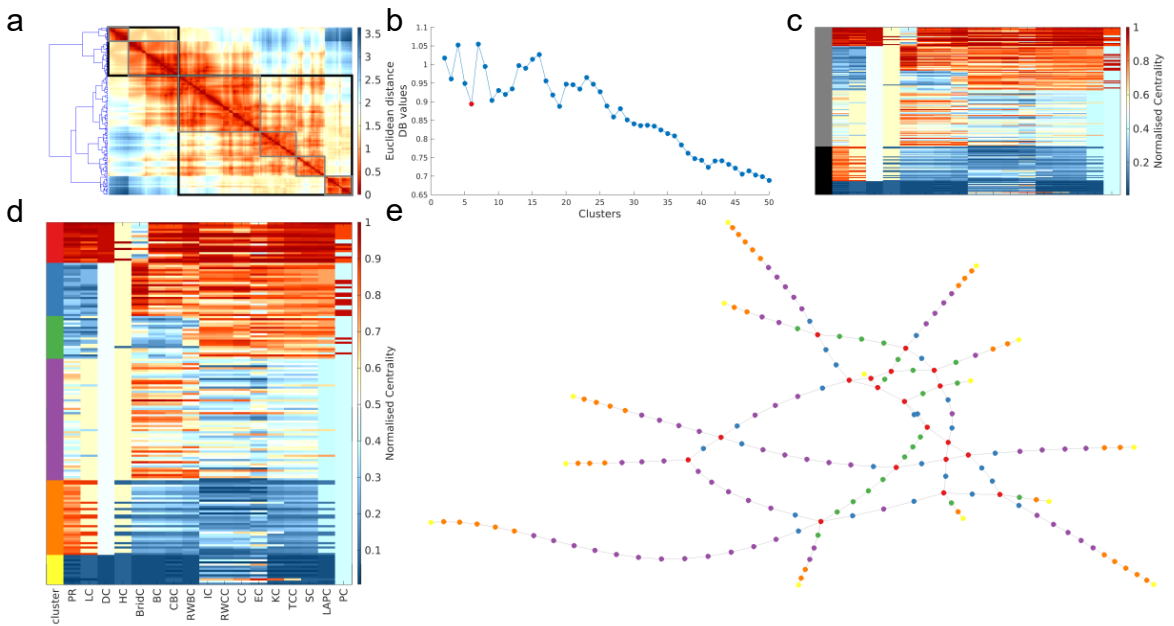

**Figure S12. Multivariate centrality profiling of the Berlin subway network.** (a) shows the dendrogram projected alongside the distance matrix of node pairs (ranks scores were normalised to be in the range 0-1 with 1 indicating the highest rank). The black and grey boxes and indicate the clusters when a two-cluster and six-cluster solution is used, respectively. (b) displays the results for the Davies-Bouldin (DB) criterion. A lower DB value represents a better clustering solution. The solution shown in (d) and (e) is labelled in red. Only the first 50 clustering solutions are shown for ease of visibility. (c) shows the matrix of nodal centrality scores (each row is a node and each column is a measure) and how these are clustered in a two-cluster solution (the black and grey represent the two different clusters). (d) shows the matrix of nodal centrality scores as well as the clusters each node was assigned to. (e) shows a topological representation of the network, produced using the force-directed layout algorithm, where each node is coloured according to the cluster it was allocated to in (d).

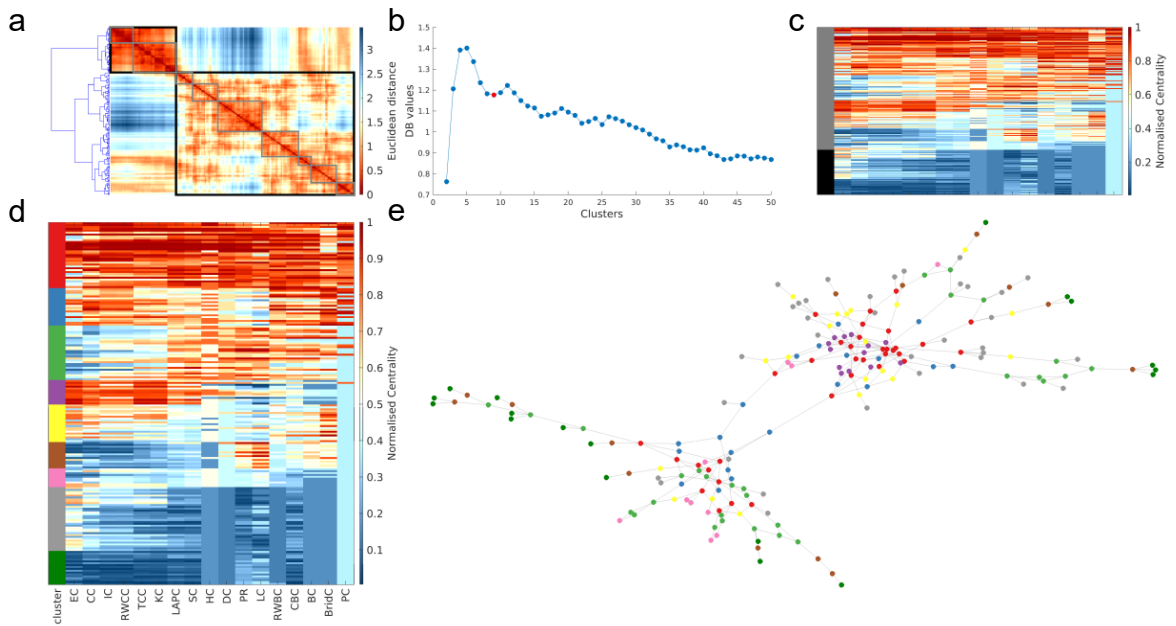

**Figure S13. Multivariate centrality profiling of the network of drug user acquaintanceships in Hartford, UK.** (a) shows the dendrogram projected alongside the distance matrix of node pairs (ranks scores were normalised to be in the range 0-1 with 1 indicating the highest rank). The black and grey boxes and indicate the clusters when a two-cluster and nine-cluster solution is used, respectively. (b) displays the results for the Davies-Bouldin (DB) criterion. A lower DB value represents a better clustering solution. The solution shown in (d) and (e) is labelled in red. Only the first 50 clustering solutions are shown for ease of visibility. (c) shows the matrix of nodal centrality scores (each row is a node and each column is a measure) and how these are clustered in a two-cluster solution (the black and grey represent the two different clusters). (d) shows the matrix of nodal centrality scores as well as the clusters each node was assigned to. (e) shows a topological representation of the network, produced using the force-directed layout algorithm, where each node is coloured according to the cluster it was allocated to in (d).

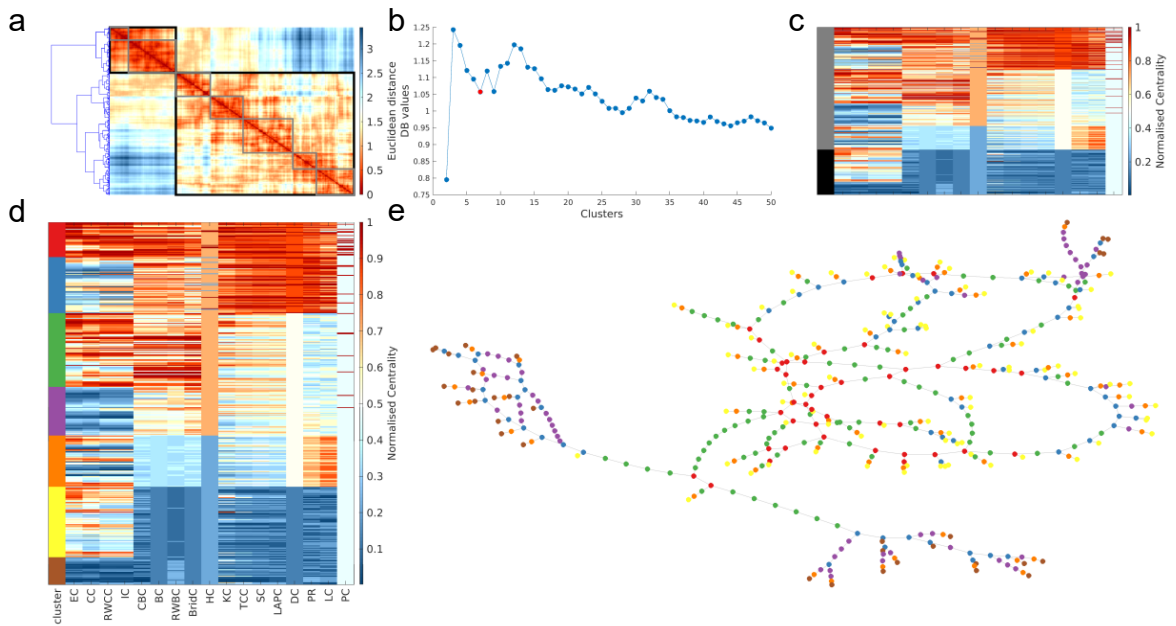

**Figure S14. Multivariate centrality profiling of the Balerma irrigation water distribution network.** (a) shows the dendrogram projected alongside the distance matrix of node pairs (ranks scores were normalised to be in the range 0-1 with 1 indicating the highest rank). The black and grey boxes and indicate the clusters when a two-cluster and seven-cluster solution is used, respectively. (b) displays the results for the Davies-Bouldin (DB) criterion. A lower DB value represents a better clustering solution. The solution shown in (d) and (e) is labelled in red. Only the first 50 clustering solutions are shown for ease of visibility. (c) shows the matrix of nodal centrality scores (each row is a node and each column is a measure) and how these are clustered in a two-cluster solution (the black and grey represent the two different clusters). (d) shows the matrix of nodal centrality scores as well as the clusters each node was assigned to. (e) shows a topological representation of the network, produced using the force-directed layout algorithm, where each node is coloured according to the cluster it was allocated to in (d).

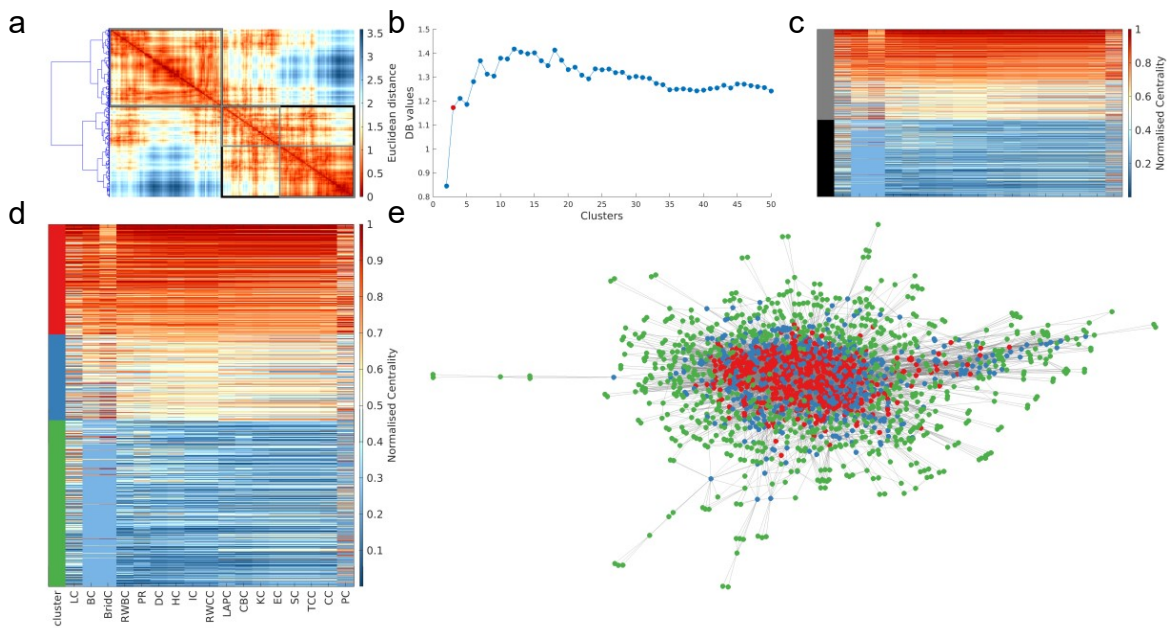

**Figure S15. Multivariate centrality profiling of the network of noun phrases (places and names) in the King James Version of the Bible.** (a) shows the dendrogram projected alongside the distance matrix of node pairs (ranks scores were normalised to be in the range 0-1 with 1 indicating the highest rank). The black and grey boxes and indicate the clusters when a two-cluster and three-cluster solution is used, respectively. (b) displays the results for the Davies-Bouldin (DB) criterion. A lower DB value represents a better clustering solution. The solution shown in (d) and (e) is labelled in red. Only the first 50 clustering solutions are shown for ease of visibility. (c) shows the matrix of nodal centrality scores (each row is a node and each column is a measure) and how these are clustered in a two-cluster solution (the black and grey represent the two different clusters). (d) shows the matrix of nodal centrality scores as well as the clusters each node was assigned to. (e) shows a topological representation of the network, produced using the force-directed layout algorithm, where each node is coloured according to the cluster it was allocated to in (d).

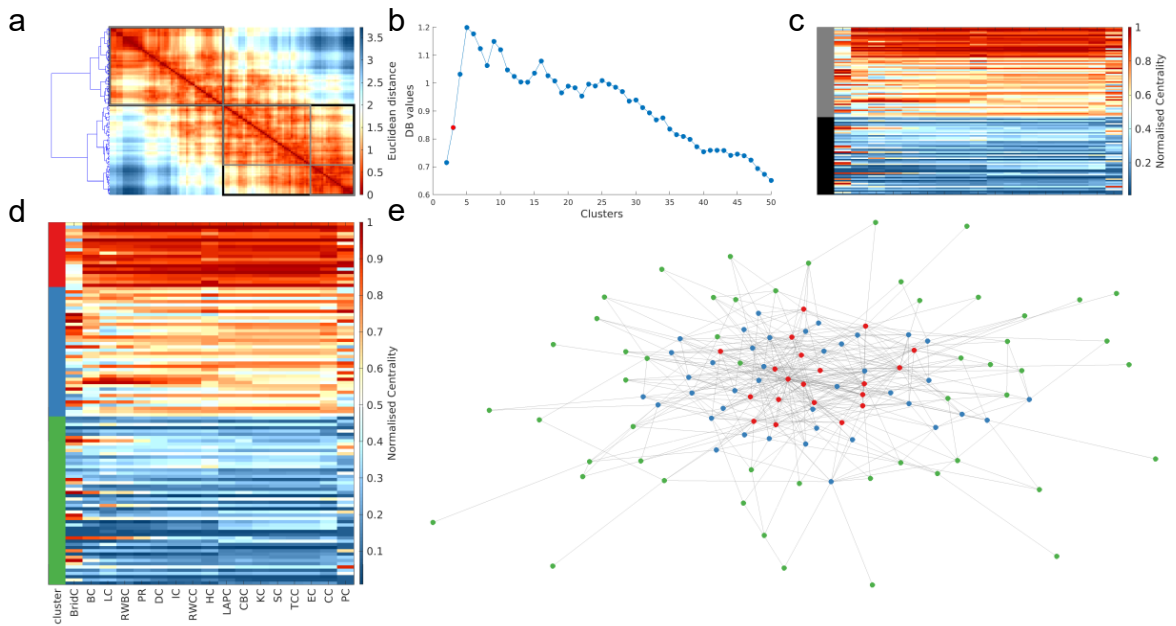

**Figure S16. Multivariate centrality profiling of the network of adjacent adjective and nouns in David Copperfield.** (a) shows the dendrogram projected alongside the distance matrix of node pairs (ranks scores were normalised to be in the range 0-1 with 1 indicating the highest rank). The black and grey boxes and indicate the clusters when a two-cluster and three-cluster solution is used, respectively. (b) displays the results for the Davies-Bouldin (DB) criterion. A lower DB value represents a better clustering solution. The solution shown in (d) and (e) is labelled in red. Only the first 50 clustering solutions are shown for ease of visibility. (c) shows the matrix of nodal centrality scores (each row is a node and each column is a measure) and how these are clustered in a two-cluster solution (the black and grey represent the two different clusters). (d) shows the matrix of nodal centrality scores as well as the clusters each node was assigned to. (e) shows a topological representation of the network, produced using the force-directed layout algorithm, where each node is coloured according to the cluster it was allocated to in (d).

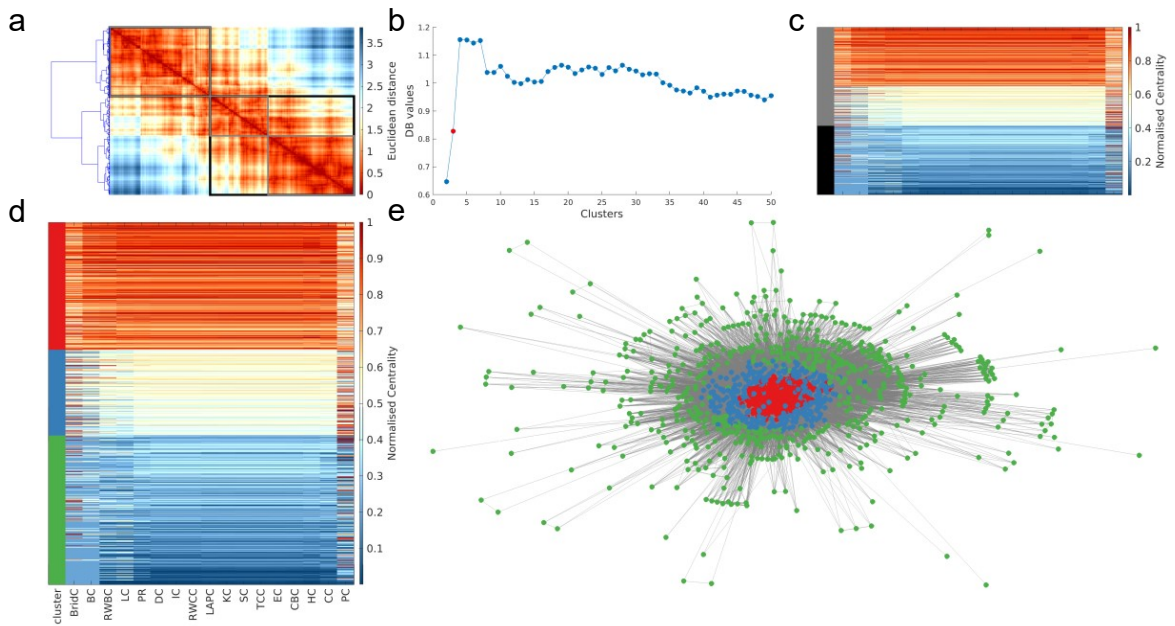

**Figure S17. Multivariate centrality profiling of the network of food ingredients and flavours.** (a) shows the dendrogram projected alongside the distance matrix of node pairs (ranks scores were normalised to be in the range 0-1 with 1 indicating the highest rank). The black and grey boxes and indicate the clusters when a two-cluster and three-cluster solution is used, respectively. (b) displays the results for the Davies-Bouldin (DB) criterion. A lower DB value represents a better clustering solution. The solution shown in (d) and (e) is labelled in red. Only the first 50 clustering solutions are shown for ease of visibility. (c) shows the matrix of nodal centrality scores (each row is a node and each column is a measure) and how these are clustered in a two-cluster solution (the black and grey represent the two different clusters). (d) shows the matrix of nodal centrality scores as well as the clusters each node was assigned to. (e) shows a topological representation of the network, produced using the force-directed layout algorithm, where each node is coloured according to the cluster it was allocated to in (d).